\documentclass[fleqn,usenatbib]{mnras}

\usepackage{newtxtext,newtxmath}
\usepackage{pdflscape}
\usepackage{newtxtext,newtxmath}
\usepackage{libertinus}
\usepackage{fix-cm}
\usepackage[T1]{fontenc}

\DeclareRobustCommand{\VAN}[3]{#2}
\let\VANthebibliography\thebibliography
\def\thebibliography{\DeclareRobustCommand{\VAN}[3]{##3}\VANthebibliography}

\usepackage{graphicx}	
\usepackage{amsmath}	
\usepackage{natbib} 

\usepackage{comment}
\usepackage{soul}

\newcommand\myphantom{\vphantom{\sqrt{2^2}}}

\title[Multiwavelength afterglow modelling]{Modelling multiwavelength afterglows of the VHE-GRB population}

\author[M. Barnard et al.]{
Monica Barnard,$^{1,2}$\thanks{E-mail: Monica.Barnard@nwu.ac.za}
Ankur Ghosh,$^{1}$\thanks{E-mail: ghosh.ankur1994@gmail.com}
Jagdish C. Joshi,$^{3,1}$ 
and Soebur Razzaque,$^{1,4,5}$\thanks{E-mail: srazzaque@uj.ac.za}
\\
$^{1}$Centre for Astro-Particle Physics (CAPP) and Department of Physics, University of Johannesburg, PO Box 524, Auckland Park 2006, South Africa\\
$^{2}$Centre for Space Research, North-West University, Potchefstroom 2520, South Africa \\
$^{3}$Aryabhatta Research Institute of Observational Sciences, Manora Peak, Nainital 263129, India\\
$^{4}$Department of Physics, The George Washington University, Washington, DC 20052, USA \\
$^{5}$National Institute for Theoretical and Computational Sciences (NITheCS), South Africa \\ 
}

\date{Accepted XXX. Received YYY; in original form ZZZ}

\pubyear{\the\year{}}

\begin{document}
\label{firstpage}
\pagerange{\pageref{firstpage}--\pageref{lastpage}}
\maketitle

\begin{abstract}
The recent detection of very high energy (VHE, $\gtrsim$ 100 GeV) $\gamma$-ray emission from gamma-ray bursts (GRBs) has provided new insights into afterglow physics. Understanding the temporal and spectral evolution of VHE GRBs requires detailed modelling of multiwavelength observations spanning radio to VHE $\gamma$ rays. Previous studies interpreted afterglow of VHE GRBs using a range of frameworks, including single- and multi-zone jet configurations, synchrotron radiation from forward and reverse shocks, synchrotron self-Compton (SSC) processes, as well as hadronic emission processes. We have modeled five long-duration VHE GRBs - GRB~180720B, GRB~190114C, GRB~190829A, GRB~201216C and GRB~221009A; using the \textsc{naima} code and modifications to it.   
The results from our analysis indicate that SSC is the dominant VHE emission mechanism, with negligible contribution from external Compton. Most VHE GRBs are well described by the forward shock model in a spherical jet configuration, where constant density interstellar medium is preferred over wind medium. Additionally, we find that VHE GRBs tend to occur in environments with lower magnetic fields and higher ambient medium densities. Interestingly, VHE GRBs lie at the edge of the $3\sigma$ region of the $E_{\rm k,iso}$ - $\epsilon_B$ correlation observed in other energetic GRBs. Our model slightly over predicts the radio fluxes, indicating that a more complicated modelling might be required in some cases. These findings provide crucial constraints on VHE GRB emission sites and mechanisms and serve as a benchmark for future observations and theoretical studies in the era of CTA and next-generation $\gamma$-ray observatories.

\end{abstract}

\begin{keywords}
(stars:) gamma-ray burst: general, radiation mechanisms: non-thermal 
\end{keywords}


\section{Introduction}
\label{sec1:intro}

The detection of very high energy (VHE, $\gtrsim$ 100 GeV) $\gamma$-rays from gamma-ray bursts (GRBs) provided us a distinct emission component to investigate particle acceleration and radiation, as well as the properties of ultra-relativistic jets and the surrounding environment. VHE photons have been detected mainly during the afterglow phase of GRBs \citep[see, e.g.,][]{2022Galax..10...66M, 2022Galax..10...74G, 2022Galax..10...67B}, and could be an extension of the late-time GeV emission~\citep[see, e.g.,][]{LATGRB_cat1_2013ApJS..209...11A, Gehrels2013FrPhy...8..661G, LATGRB_cat2_2019ApJ...878...52A}. The afterglow phase begins, following the prompt emission in MeV, as the jet interacts with the surrounding medium, producing external shocks. These shocks accelerate particles and amplify magnetic fields ($B$-fields), leading to multiwavelength radiation, analysing which can provide crucial information about energy {release}, jet structure {and} composition, and the surrounding environment of these bursts~\citep[see, e.g.,][]{piran_2004RvMP...76.1143P, Meszaros2006RPPh...69.2259M, kumar2015PhR...561}.

Currently, VHE $\gamma$-ray afterglow emission has been {detected} from six {long-duration} GRBs, including GRB~180720B \citep{Abdalla2019}, GRB~190114C \citep{2019Natur.575..455M}, GRB~190829A \citep{HESS:2021dbz}, GRB~221009A \citep{doi:10.1126/sciadv.adj2778}, GRB~201015A \citep{2020GCN.28659....1B} and GRB~201216C \citep{Abe2024}. Detection of VHE emission from a short burst, GRB~160821B, has also been reported at $3\sigma$ level \citep{Acciari2021ApJ..90A}. All these discoveries have been made with ground-based $\gamma$-ray detectors such as High Energy Stereoscopic System (H.E.S.S.), Major Atmospheric Gamma Imaging Cherenkov Telescopes (MAGIC), and Large High Altitude Air Shower Observatory (LHAASO). A natural question to ask is the emission mechanism of VHE $\gamma$ rays from these GRBs. Synchrotron radiation by shock-accelerated electrons successfully explains radio, optical, X-ray and GeV $\gamma$-ray afterglows. However, it is challenging to explain VHE $\gamma$ rays with synchrotron radiation, especially the detection of GRB~190114C in up to $\approx 1$ TeV $\gamma$ rays by MAGIC \citep{2019Natur.575..455M,2019Natur.575..459M} and of GRB~221009A in $\gtrsim 10$ TeV $\gamma$ rays by LHAASO \citep{cao_2023S_lhaaso2778C}.

The synchrotron self-Compton (SSC) has been explored to predict VHE emission from the afterglow forward shock in the past~\citep[see, e.g.,][]{dermer_50GeV_IC, Panai2000ApJ543, sari_IC_paper, Nakar:2009er}. Discovery of VHE emission led to modelling multiwavelength data from GRBs using the synchro-SSC emissions~\citep[see, e.g.,][]{Abdalla2019, 2019Natur.575..455M, wang2019ApJ.117W, derishev2021ApJ.135D, Joshi2021, fraija_2021ApJ.12F, yamasaki_2022MNRAS.2142Y, Salafia2022, 2023MNRAS.520..839K}. Compton scattering of external radiation fields, such as the cosmic microwave background (CMB) and infrared (IR) backgrounds, has also been considered for modelling the multiwavelength data together with synchrotron and SSC \citep{zhang2021ApJ..55Z, 2023ApJ...947L..14Z, Barnard2024}. Modelling the long-term multiwavelength observations of GRB~190114C by \citet{Misra2021MNRAS_5685M} found that the microphysical parameters of the synchro-SSC model may evolve with time. \citet{irene2023MNRAS_149G} studied the afterglow properties of GRB~180720B, GRB~190114C, and GRB~221009A and found that these bursts might have occurred in the low-density circumburst medium if the SSC mechanism produces the VHE radiation.

The ultra high-energy protons, if co-accelerated with the electron in GRBs, can radiate VHE synchrotron photons, because of a proton's mass heavier than an electron~\citep{Totani1998ApJ...502L..13T, Razzaque2010OAJ.....3..150R, Razzaque2010ApJ...724L.109R}. This emission mechanism has been investigated in the forward shock \citep{2023ApJ...955...70I, isra2023ApJ956_12I} and in the reverse shock \citep{2023ApJ...947L..14Z} to model VHE afterglows. Photohadronic interactions by ultra high-energy protons with afterglow photons have also been investigated to model VHE emission from GRBs~\citep{sahu2020ApJ.41S, 2022ApJ...929...70S,Klinger_2024}. 
It should be noted that the total energy requirement for these models often exceeds the observed $\gamma$-ray energy release. Ultra high-energy protons, if escaping as cosmic rays, can interact with the CMB and IR backgrounds to produce line-of-sight VHE emission to explain $\gtrsim 10$ TeV photons detected by LHAASO from GRB~221009A~\citep{Batista2022arXiv221012855A, Das:2022gon, Mirabal2023MNRAS.519L..85M}.

Multi-zone models with synchro-SSC emissions have also been proposed to model multiwavelength data of GRBs detected with VHE $\gamma$ rays. The most natural of these models is a combination of reverse- and forward-shock emissions used for modelling GRB~180720B data ~\citep{arimoto2024NatAs_134A}. Two-component jet models, i.e., combinations of a narrow and a wider jets ~\citep{2021MNRAS.504.5647S, sato2023JHEAp51S, 2023MNRAS.522L..56S}, or a core and a wing ~\citep{Ren2023,Zheng_2024} or a structured jet ~\citep{oconnor2023SciA....9I1405O} have also been used to explain the details of the afterglows in VHE GRBs. A two-zone model having different $B$-field strengths was developed for GRB~190829A, in which the weaker $B$-field electrons are scattered by the synchrotron photons available in the stronger $B$-field to produce an intensity of VHE $\gamma$-rays \citep{Khangulyan2023ApJ_87K}.

In this work, we fit multiwavelength data from the population of VHE GRBs with synchro-Compton emission to the extent of a single-zone, namely the forward shock, scenario. For this purpose, we employ the publicly-available \textsc{naima} code~\citep{2015ICRC...34..922Z}~\footnote{Additional documentation on the code and the installation thereof is available at \url{http://naima.readthedocs.org}.} with some modifications. Our goal is to exhaustively explore this one-zone scenario by fitting as much broadband and multi-epoch data as possible for the VHE GRBs. This work, therefore, can identify cases where models with multi-emission zones are necessary.

The outline of this paper is the following. In Section~\ref{sec2:model}, we {briefly} discuss the afterglow emission model used in our work. In Section~\ref{sec3:data}, we describe the data for each GRB that was analysed and used {for fitting}. In Section~\ref{sec4:multi_sed}, we apply the \textsc{naima} code, described in Appendix~\ref{sec:appA_naima}, to the aforementioned GRBs and compare our results to the multiwavelength observations, with the discussion following in Section~\ref{sec5:concl}. Lastly, the blastwave modelling for different scenarios in the \textsc{naima} code and the Markov chain Monte Carlo (MCMC) technique are described in detail in Appendices~\ref{sec:appA_naima} and ~\ref{sec:appB_MCMC}.

\section{Afterglow emission modelling}
\label{sec2:model}

We perform afterglow modelling in this work with emission from the forward shock of the GRB blastwave after the deceleration time ($t_{\rm dec}$), when it evolves following the self-similar set of equations in either a constant-density inter-stellar medium (ISM) or in a pre-existing stellar wind of the progenitor star, with density $\propto R^{-2}$ where $R$ is the radius from the star \citep{Blandford1976}. More details on how the evolution of the forward-shock radius and bulk Lorentz factor are implemented in the \textsc{naima} code are given in Appendix~\ref{sec:blastwave}. {In modelling the GRB afterglow emission, \textsc{naima} uses a steady-state approximation for the underlying particle distributions and radiative processes. This approach assumes that the electron population reaches equilibrium on timescales shorter than or comparable to the dynamical evolution of the blast wave. This simplification is commonly used and captures the general spectral and temporal behaviour of GRB afterglows. We are aware that the steady-state approximation may introduce a systematic uncertainty in the parameter estimates due to the time-dependent nature of the GRB afterglow environment. However, this type of model uncertainty is difficult to quantify and has not been explicitly accounted for in the present analysis. As a result, the reported uncertainties reflect only the statistical errors, and the total error budget may be underestimated.} Note also that \textsc{naima} does not calculate $t_{\rm dec}$ or produce light curves, and thus we use Eqs.~[1] and [2] from \citet{Barnard2024} to estimate the initial Bulk Lorentz factor $\Gamma_0$ of the blastwave. We set the $t_{\rm dec}\equiv t_{p}$ (peak time) or $\equiv T_{90}$ (the time duration for a burst to emit from 5\% to 95\% of its total measured counts), by studying the afterglow light curves of previous studies, and depending on the longer time duration at which the afterglow emission is expected to occur. The values of $t_{p}$ and $T_{90}$ are associated with the lab frame, thus we apply a factor $(1+z)$ correction, with $z$ the redshift, to convert these times in the rest frame of the burst. In the wind scenario $A=3.02\times10^{35}A_*$ cm$^{-1}$, with $A_*\equiv\dot{M}_{\rm w}/v_{\rm w}=\dot{M}_{-5}/v_{8}$ recorded in Table~\ref{tab:bestfit_params} along with the ISM density $n$ in cm$^{-3}$ (see \citet{Barnard2024} for more details).

\textsc{naima} has been developed for modelling of galactic objects, however it has also {been} extended for the SSC modelling of GRBs.\footnote{https://github.com/Carlor87/GRBmodelling} In this code the synchrotron radiation is calculated using the formalism as described by \citet{2010PhRvD..82d3002A} while the inverse Compton (IC) radiation with non-thermal and thermal photons is based on \citet{1981ApnSS..79..321A, 2014ApJ...783..100K}. The details of synchrotron, SSC and external Compton (EC) emission calculations in \textsc{naima} are discussed in Appendices \ref{sec:synchrotron}, \ref{sec:ssc} and \ref{sec:ec}, respectively. The modification we have made to calculate the maximum electron energy is described in Appendix \ref{sec:modify}.

In addition to the calculation of emissions, \textsc{naima} also calculates $\gamma\gamma$ attenuation within the emission region. The opacity for the $\gamma\gamma$ attenuation calculation within the emission region follows \citet{Aharonian2004} and \citet{Eungwanichayapant2009}. The details of the $\gamma\gamma$ attenuation are described in Appendix \ref{sec:gamma_gamma_abs}. \textsc{naima} includes a function to calculate the loss of energy due to the interaction of $\gamma$-rays with the EBL, although we followed the approach used by \citet{Barnard2024} and computed the EBL attenuation using the model of \citet{dominguez_extragalactic_2011}. 

We have employed the MCMC technique \citep{ForemanMackey2013} available within \textsc{naima} to fit multiwavelength afterglow spectral energy distribution (SED) for each GRB during a time interval with the most data coverage. The best-fit model parameters thus obtained are then used to calculate model SEDs in other time intervals. The details of the fitting procedure with initialisation of the models, priors on the parameters, and selection process for the model are described in Appendices \ref{sec:initparams} and \ref{sec:fitting_selection}, respectively.

\section{Data acquisition and analysis}
\label{sec3:data}

We have collected radio, optical and $\gamma$-ray (MeV-VHE) data for our GRB sample from the published papers, while we have analysed X-ray data from \textit{Swift}-XRT by ourselves. We describe this analysis and the data sets used for modelling in this section.

\subsection{\textit{Swift}-XRT data analysis}
\label{subsec:XRT}
The \textit{Swift}-XRT observations of GRBs are automatically processed by the UK \textit{Swift} Science Data Center (UKSSDC), which generates light curves, time-averaged spectra, and time-sliced spectra.\footnote{\url{https://www.swift.ac.uk/}} For this study, we focus on the time-sliced spectra to match the time-stamp of the multiwavelength data for the {VHE} GRBs. Both photon counting (PC) mode and window timing (WT) mode data were considered, ensuring comprehensive temporal coverage. The time-sliced XRT spectra were selected only for intervals coincident with observations in other wavelengths, such as optical, infrared, and radio, to generate precise, multi-epoch SEDs.

The XRT time-sliced spectra were analysed using the \textsc{xspec} software (version 12.14.0b) within the \textsc{heasoft} package, following standard reduction procedures. To improve the reliability of the spectral analysis, we excluded energy channels below 0.5 keV and above 10 keV, as these are often affected by increased uncertainties and instrumental limitations. The initial steps in the spectral fitting involved loading the source spectra, background spectra, and associated response matrix files (RMFs) and ancillary response files (ARFs). We modelled the XRT spectra using an absorbed power-law model, incorporating both Galactic and intrinsic absorption. For {the} Galactic absorption, we used the \textit{tbabs} model \citep{2000Wilms}, which accounts for absorption due to the Galactic ISM with fixed hydrogen column density values. For the intrinsic absorption (i.e., absorption occurring within the host galaxy of the GRB), we employed the \textit{ztbabs} model \citep{2000Wilms}, which allows for redshift-dependent absorption.

Post-fitting, the photon flux and photon index ($\Gamma_i$) were extracted from the best-fit model. These parameters were then used to calculate the unabsorbed energy flux by integrating the flux over the XRT energy range (0.5 – 10~keV). The energy-averaged flux was calculated with the corresponding $1\sigma$ error using \textsc{xspec}.

\subsection{Multiwavelength afterglow SEDs}
\label{sec4:multi_sed}
{We describe multiwavelength SED fitting results for each VHE GRB in this section.}
{See {Appendix}~\ref{sec:appB_MCMC} 
for a detailed discussion of how we have used \textsc{naima} to fit and select the best-fit model to the multiwavelength observations, as well as the diagnostic plots for each GRB studied in this paper.}
{The results of the best-fit parameters for each GRB in the case of ISM and wind are compiled in Tables~\ref{tab:bestfit_params} and ~\ref{tab:BIC_values}.} 
Since \textsc{naima} does not include a reverse shock component, which could be important during the very early afterglow phase, we do not use data from the epochs that are too close to the prompt emission. Furthermore, we only use spectral data available in the literature for our modelling, and not the light curve data. This is because conversion of data from light curves to spectra is non-trivial.

\subsubsection{{GRB~180720B}}
\label{sec:grb180720B}

\begin{figure*}
    \centering
    \includegraphics[width=0.45\textwidth]{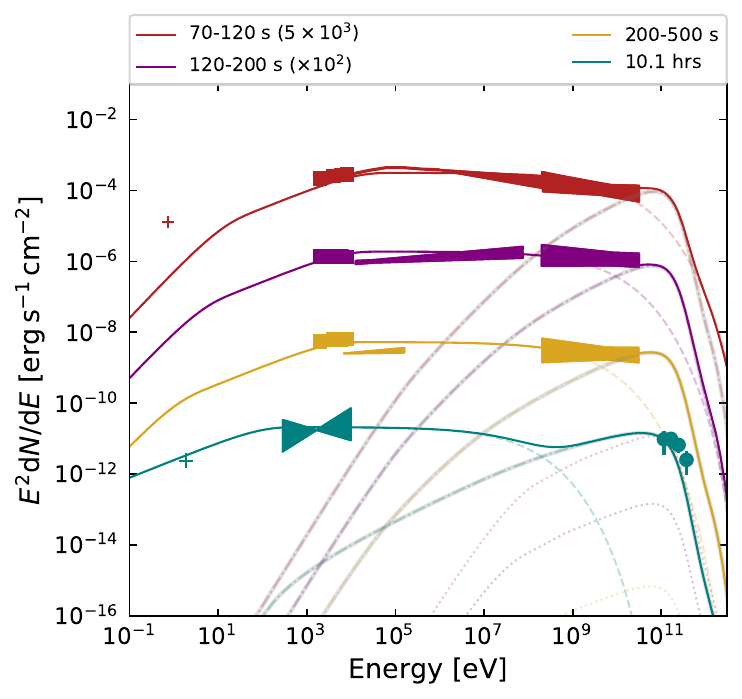}
    \hspace{0.3cm}
	\includegraphics[width=0.45\textwidth]{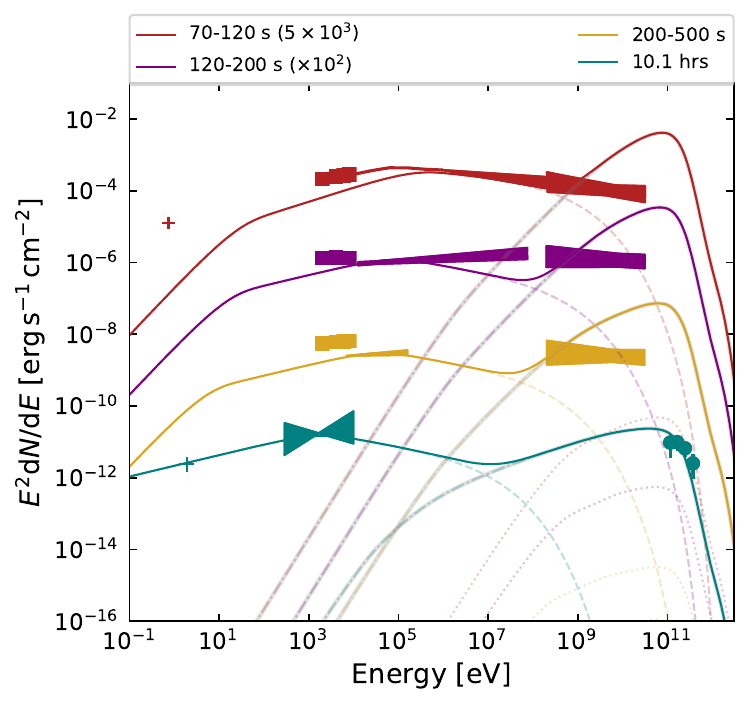}
	\caption{The SEDs for GRB~180720B at several epochs, as indicated in the legend, for the ISM (left) and wind (right) density scenarios. For each epoch the synchrotron (dashed), SSC (dashed dotted), EC (dotted) and total (solid) emission are modelled and fitted to the multiwavelength data, i.e., optical (crosses), \emph{Swift}-XRT (squares), \emph{Swift}-BAT and \emph{Fermi}-LAT (shaded regions), and H.E.S.S. (circles). {For the 10~hr epoch the X-ray data is an extrapolation to the time window of the H.E.S.S. observation. The data were collected from} \citet{2018GCN.22977....1S,Abdalla2019,Fraija_2019,Ronchi2020} and \url{https://www.swift.ac.uk/}.}
    \label{fig:sed_180720b}
\end{figure*}

VHE emission from GRB~180720B was detected approximately ten hours after the initial trigger {by the H.E.S.S. telescope} \citep{Abdalla2019}. This is the only GRB where much-delayed VHE observation was evident. For this epoch the optical and VHE data were sourced from \citet{Abdalla2019,Fraija_2019}, with the X-ray data an extrapolation to the time window of the H.E.S.S. observation. We obtained the optical, soft X-ray to $\gamma$-ray data for the other epochs from \citet{Ronchi2020, 2018GCN.22977....1S}. Multiwavelength SEDs have been constructed for four distinct observational epochs. We have not selected data from very early epochs of GRB~180720B that are too close to the prompt emission. We perform a Bayesian information criterion (BIC) using MCMC for the epoch at 10 hours post-burst when H.E.S.S. data was available, for both ISM and wind models. The combination of synchrotron, SSC, and EC is taken into consideration to model the multiwavelength data and the EC contribution was found to be negligible. Based on the SED fitting and corner plots as shown in Figs. \ref{fig:mcmc_180720b_ism}, \ref{fig:corner_180720b_ism}, \ref{fig:mcmc_180720b_wind}, and \ref{fig:corner_180720b_wind}, as well as the multi-epoch SED shown in Fig.~\ref{fig:sed_180720b}, we found the ISM model as the better fit. The best fit values of model parameters and BIC numbers for both the mediums are tabulated in Tables \ref{tab:bestfit_params} and \ref{tab:BIC_values}, respectively. We translated the value of the $B$-field for other epochs, by calculating the ratio between a given $B$-field energy and shock energy $\epsilon_B$ given in Eq.~(\ref{eq:epsB}), by keeping other model parameters fixed. 

At a $z=0.653$ \citep{2018GCN.22996....1V}, GRB~180720B is a {highly energetic} event, with an isotropic-equivalent energy release of $E_{\rm k,iso} \approx 6 \times 10^{53}$ and $1 \times 10^{54}$ erg for the ISM and wind cases,  respectively. The $\Gamma_0$, {using $t_{\rm dec}\equiv T_{p}=78$~s \citep{Ronchi2020} is larger for the ISM case}. GRB~180720B displays afterglow characteristics closely aligned with {those} of GRB~190114C, except having low number density for the ISM medium, indicating possible similarities between the burst environment and dynamics.

\subsubsection{{GRB~190114C}}
\label{sec:grb190114C}

\begin{figure*}
    \centering
	\includegraphics[width=0.45\textwidth]{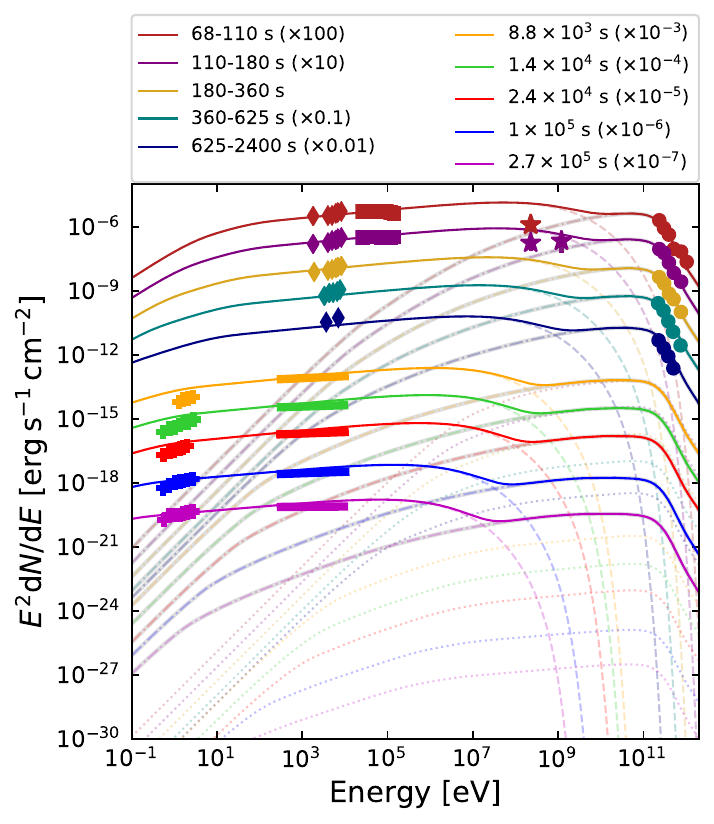}\hspace{0.3cm}
	\includegraphics[width=0.45\textwidth]{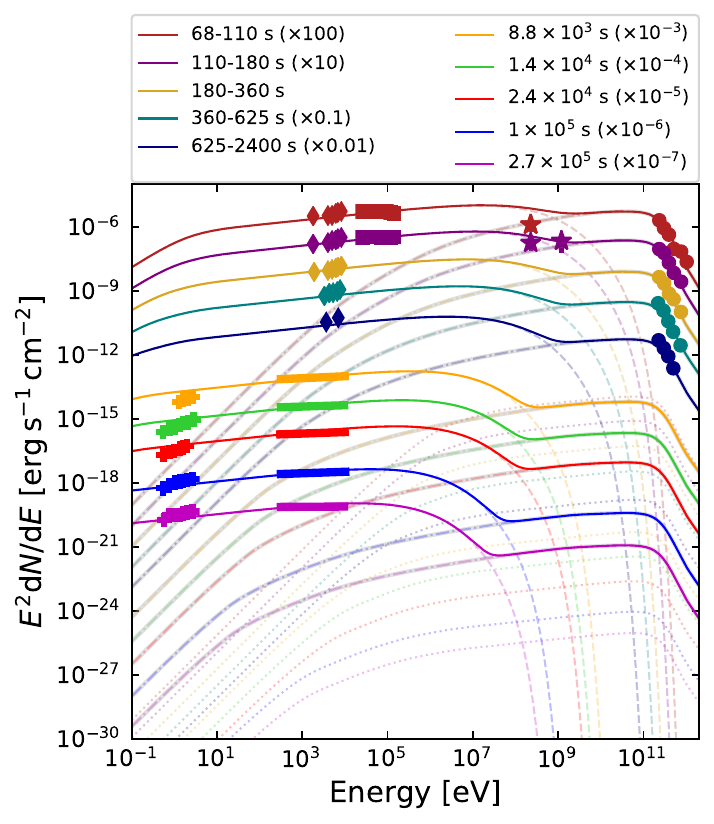}
    \caption{The SEDs for GRB 190114C at several epochs, as indicated in the legend, for both the ISM (left) and wind (right) density scenarios. For each epoch the synchrotron (dashed), SSC (dashed dotted), EC (dotted) and total (solid) emission are modelled and fitted to the multiwavelength data, i.e., optical (crosses), \emph{Swift}-XRT (diamonds and shaded regions), \emph{Swift}-BAT (squares), Fermi-LAT (stars), and MAGIC (circles). The data were collected from \citet{2019Natur.575..459M} and \url{https://www.swift.ac.uk/}.}
    \label{fig:sed_190114c}
\end{figure*}

GRB~190114C was the first GRB to have VHE photons detected by the MAGIC telescope, with the detection occurring around 60 seconds post-trigger \citep{2019Natur.575..459M}. VHE data for GRB~190114C was collected for five different epochs, spanning from 68 seconds to 2400 seconds after the trigger. Furthermore, we also acquired late-time optical and X-ray data from \citet{2019Natur.575..459M} to obtain better constraint on the models. In case of GRB~190114C, we have not included late-time radio or optical data from \citet{Misra2021MNRAS_5685M} in our modelling because those data are given only in light curves. The multi-epoch SEDs of GRB~190114C are shown in Fig.~\ref{fig:sed_190114c} for ISM and wind mediums, respectively. {A} total of 10 epochs {were used} for the SED modelling. Among them are five time bins that have VHE data, and the rest have late-time optical to X-ray data coverage. {The} VHE data presented in the study are the observed data before internal and EBL absorption corrections. We selected the second epoch (110-180 s) to run the MCMC sampler because of its good data coverage from X-ray to VHE. After fitting, we obtain our best fit parameters for both the ISM and wind mediums, as tabulated in Table~\ref{tab:bestfit_params} and the posterior distributions are shown in Figs.~\ref{fig:corner_190114c_ism} and \ref{fig:corner_190114c_wind}. Among the model parameters, the $B$-field is the only time-dependent parameter. Similar techniques are used to translate parameters for all the epochs as mentioned in Section~\ref{sec:grb180720B} for GRB~180720B. As GRB~190114C occurred at $z = 0.45$ \citep{2019GCN.23695....1S, 2019GCN.23708....1C}, {flux attenuation due to the} EBL is much higher than {the internal} $\gamma\gamma$ absorption.

From Table \ref{tab:bestfit_params}, it is evident that GRB~190114C is very energetic ($E_{\rm k,iso} = 8 \times 10^{53}$ erg) and occurred in a denser ISM ($n_0 = 5$~cm$^{-3}$). We calculated $\Gamma_0 = 167$ and 107 for the ISM and wind, respectively using $t_{\rm dec}\equiv T_{90}=116$~s \citep{Ajello_2020}. The optical to VHE data from early to later epochs are well fitted with both the ISM and wind models. From Figs.~\ref{fig:mcmc_190114c_ism}, \ref{fig:corner_190114c_ism}, \ref{fig:mcmc_190114c_wind}, \ref{fig:corner_190114c_wind}, and the composite SED shown in Fig. \ref{fig:sed_190114c}, it is hard to distinguish which model is preferred. However, the BIC values recorded in Table~\ref{tab:BIC_values} clearly indicates preference for the wind case, even though the high-energy cutoff of the electron distribution $E_c$ is not properly constrained.

\subsubsection{{GRB~190829A}}
\label{sec:grb190829A}

\begin{figure*}
    \centering
	\includegraphics[width=0.45\textwidth]{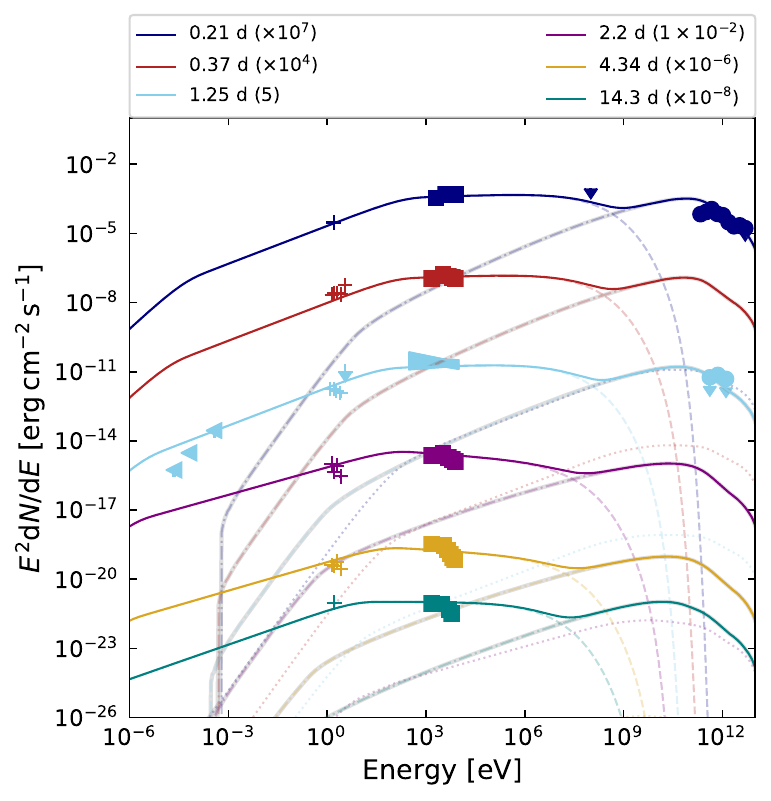}
    \hspace{0.3cm}
	\includegraphics[width=0.45\textwidth]{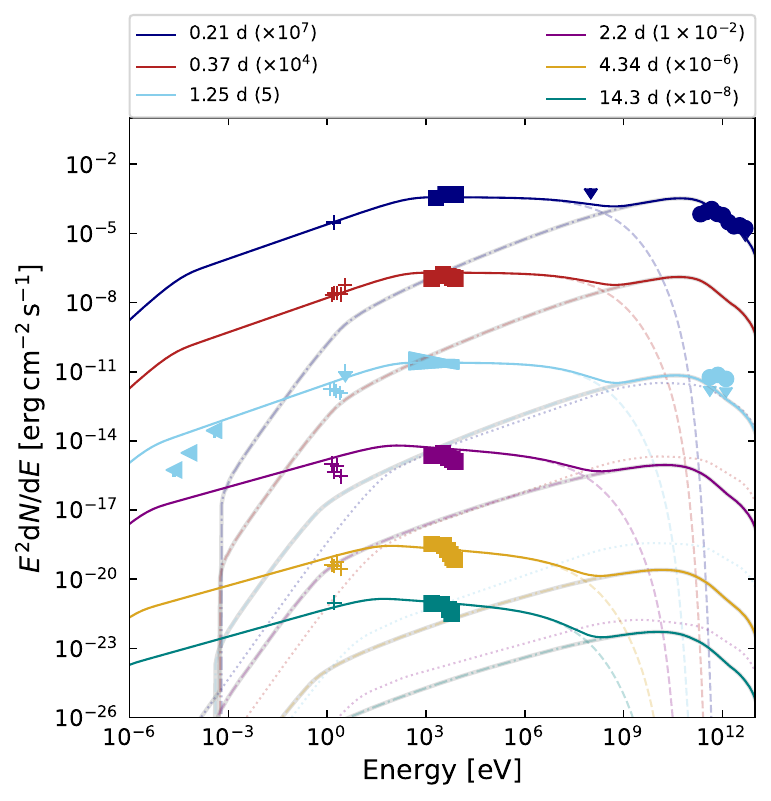}
    \caption{The SEDs for GRB 190829A at several epochs, as indicated in the legend, for the ISM (left) and wind (right) density scenarios. For each epoch the synchrotron (dashed), SSC (dashed dotted), EC (dotted) and total (solid) emission are modelled and fitted to the multiwavelength data, i.e., radio (triangles), optical (crosses), \emph{Swift}-XRT (squares and shaded region), \emph{Fermi}-LAT upper limit, and {H.E.S.S.} (circles). The data were collected from \citet{2019GCN.25589....1D,2019GCN.25676....1L,2020MNRAS.496.3326R,HESS:2021dbz,Hu2021} and \url{https://www.swift.ac.uk/}.}
    \label{fig:sed_190829a}
\end{figure*}

GRB~190829A, the lowest luminosity VHE detected GRB to date, occurred at a relatively lower $z = 0.0785$ \citep{2019GCN.25565....1V}. VHE data for two distinct epochs, corresponding to 5 and 30 hours (i.e., 0.21 and 1.25 days, respectively) after the GRB trigger, were extracted from \citet{HESS:2021dbz}, as illustrated in Fig. \ref{fig:sed_190829a}. The optical and radio observations for these two epochs were obtained from \citet{2019GCN.25589....1D,2019GCN.25676....1L,2020MNRAS.496.3326R,Hu2021,Salafia2022}, while the optical and radio data for the remaining epochs were sourced from \citet{Hu2021}. The SED of GRB~190829A was constructed across multiple epochs from 0.21-14.3 d since the burst, whereas the MCMC sampling was performed for the first epoch (0.21 d) because of its robust and simultaneous data coverage across multiple wavelengths. Same models and mediums were used to fit GRB~190829A data like {the} other two GRBs. 
Due to the proximity of GRB~190829A, the absorption in EBL was significantly lower than that seen in other VHE GRBs. The ISM model is the more suitable fit over wind because the wind model under-predicts the XRT data as shown in Figs.~\ref{fig:mcmc_190829a_ism} and \ref{fig:mcmc_190829a_wind}. The best-fit parameters for the ISM and wind scenarios, along with the corresponding posterior distributions, are presented in Table \ref{tab:bestfit_params} and \ref{tab:BIC_values}, and in Figs. \ref{fig:corner_190829a_ism} and \ref{fig:corner_190829a_wind}. For the subsequent epochs, the SED fitting was repeated by adjusting only the $B$-field strength while keeping other parameters fixed. 

GRB~190829A stands out as a unique VHE-detected GRB due to its low $E_{\rm k,iso} = 1 \times 10^{52}$ erg and higher $n_0=30$~cm$^{-3}$. Furthermore, GRB~190829A has low $\Gamma_0$ values using $t_{\rm dec}\equiv T_{90}=63$~s \citep{Lesage2019} for the ISM and wind cases, respectively, compared to the other VHE GRBs.

\subsubsection{GRB~201216C} 
\label{sec:grb201216c}

\begin{figure*}
    \centering
    \includegraphics[width=0.45\textwidth]{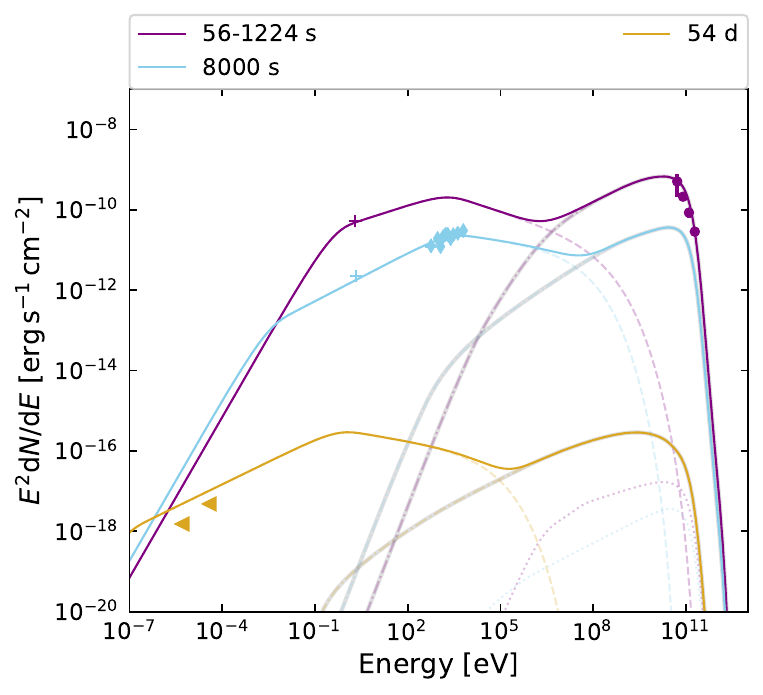}
    \hspace{0.3cm}
	\includegraphics[width=0.45\textwidth]{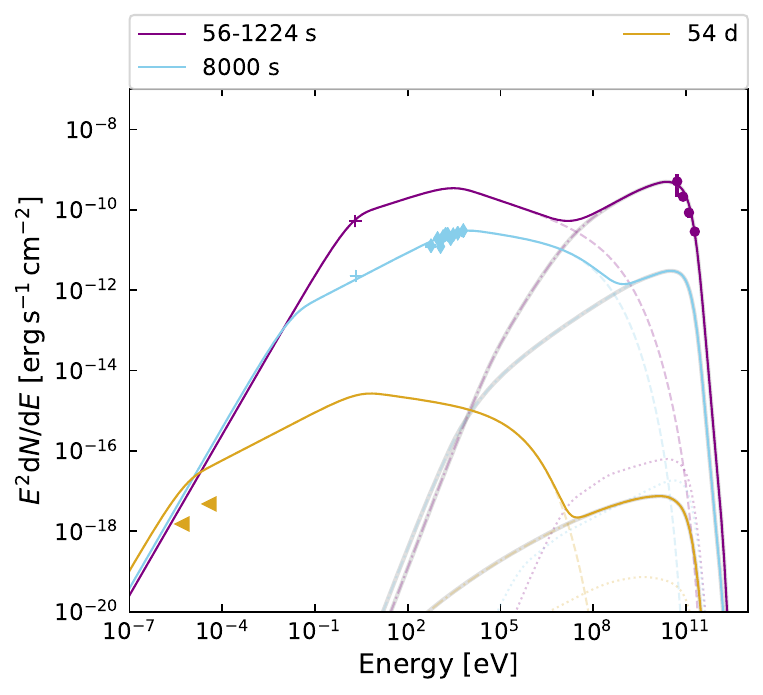}
    \caption{The SEDs for GRB 201216C at several epochs, as indicated in the legend, for the ISM density (left) and the stellar wind (right) scenarios. For each epoch the synchrotron (dashed), SSC (dashed dotted), EC (dotted) and total (solid) emission are modelled and fitted to the multiwavelength data, i.e., radio (triangles), optical (crosses), \emph{Swift}-XRT (diamonds), and MAGIC (circles). The data were collected from \citet{2020GCN.29066....1I,2022MNRAS.513.1895R} and \citet{Abe2024}.}
    \label{fig:sed_201216c}
\end{figure*}

At a $z = 1.1$ \citep{2020GCN.29066....1I}, GRB~201216C is {the} most distant GRB to date for which VHE photons were detected by MAGIC in between 56 and 1200 s after the \textit{Swift}-BAT trigger \citep{Abe2024}. Simultaneous detection at other wavelengths is unavailable for the latter epoch, but for the other epochs data were acquired from \citet{2020GCN.29066....1I, 2022MNRAS.513.1895R} and \citet{Abe2024}. So we performed MCMC sampling at 56-1224 s SED, where VHE data points are available. Like other GRBs, the contribution of the EC is negligible compared to the SSC for GRB~201216C. The best-fit SEDs and the posterior distributions for both ISM and wind-like environments are presented in Figs.~\ref{fig:mcmc_201216c_ism}, \ref{fig:corner_201216c_ism}, \ref{fig:mcmc_201216c_wind} and \ref{fig:corner_201216c_wind}. These results do not allow for a clear distinction between the two circumburst medium scenarios except for the BIC value being lower for the ISM case. The $E_{\rm k,iso} = 6 \times 10^{53}$ erg for both scenarios, although $\Gamma_0$ is roughly four times higher for the ISM case, using $t_{\rm dec}\equiv T_{p}=60$~s \citep{Abe2024} for both cases respectively. The corresponding best-fit parameters are summarised in Table~\ref{tab:bestfit_params} and \ref{tab:BIC_values}.

GRB~201216C shows similar characteristics like those of GRB~190114C and GRB~180720B. Our model over-predicts the late-time radio data for GRB~201216C in both the ISM and wind scenarios.

\subsubsection{GRB~221009A}
\label{sec:grb221009a}

\begin{figure*}
    \centering
    \includegraphics[width=0.45\textwidth]{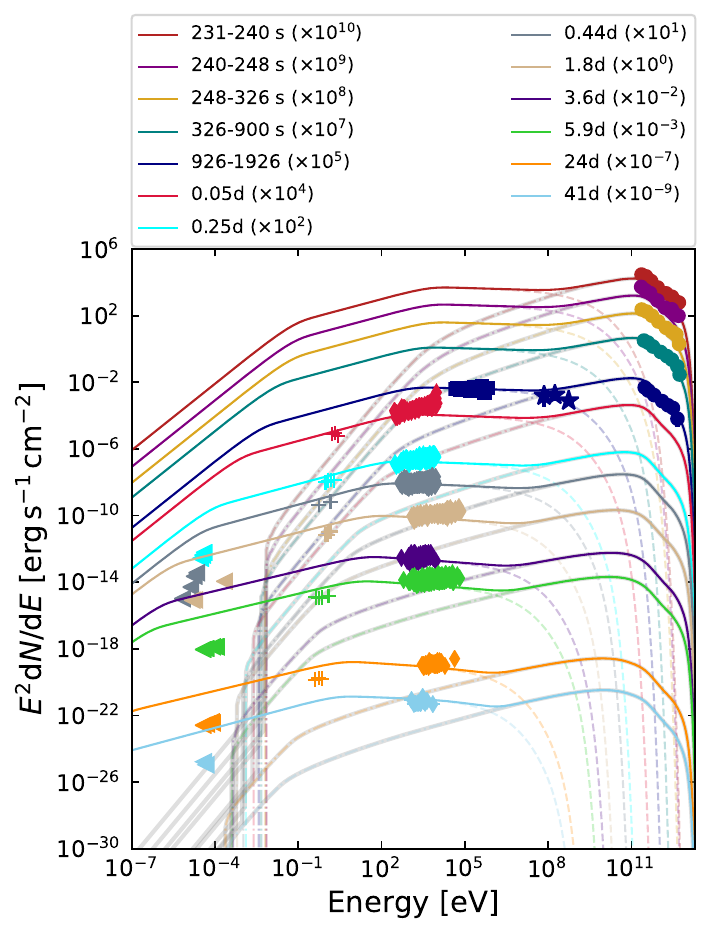}
    \hspace{0.3cm}
	\includegraphics[width=0.45\textwidth]{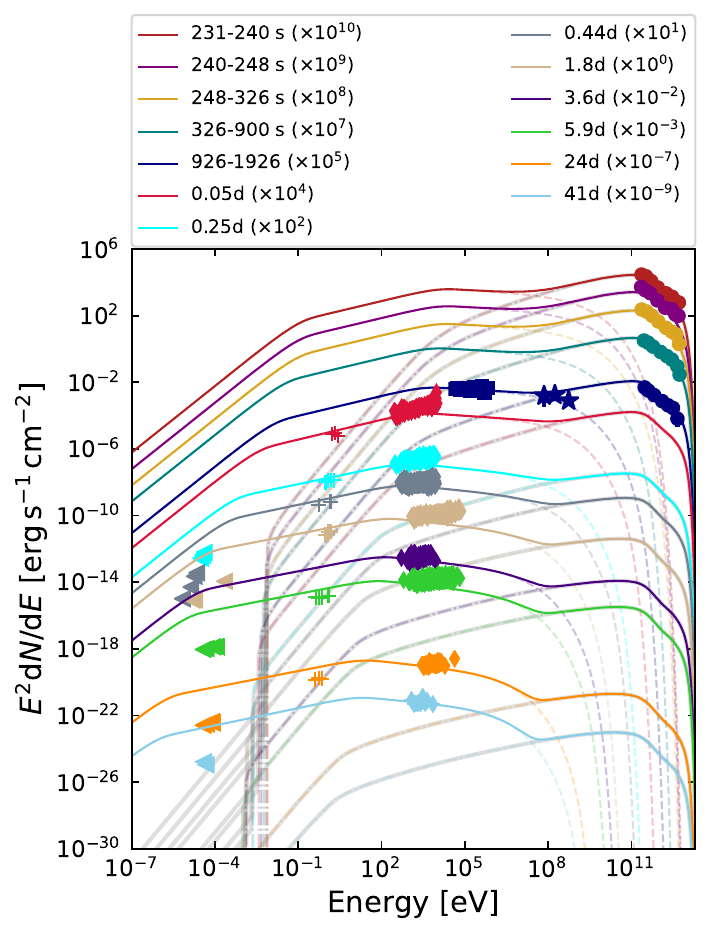}
    \caption{The SEDs for GRB 221009A at several epochs, as indicated in the legend, for the ISM density (left) and the stellar wind (right) scenarios. For each epoch the synchrotron (dashed), SSC (dashed dotted), and total (solid) emission are modelled and fitted to the multiwavelength data, i.e., radio (triangles), optical (crosses), \emph{Swift}-XRT (diamonds), {GBM (squares), \emph{AGILE} (stars)}, and LHAASO (circles). The data were collected from \citet{2023NatAs...7..986B,2023Sci...380.1390L,oconnor2023SciA....9I1405O,2023ApJ...956L..23T} and \citet{Banerjee2024}. The EC was omitted for figure clarity.}
    \label{fig:sed_221009a}
\end{figure*}

GRB~221009A at a $z = 0.151$ \citep{2022GCN.32648....1D, 2022GCN.32686....1C}, is recognised as the brightest GRB ever recorded \citep{Burns2023ApJ...946L..31B}. It exhibited exceptionally high-energy emission, with the detection of {up to} 18 TeV photons reported by LHASSO \citep{Huang2022}. {VHE data for five distinct epochs were obtained from \citet{2023Sci...380.1390L}, while additional multiwavelength datasets across various frequencies were compiled from both \citet{2023NatAs...7..986B,oconnor2023SciA....9I1405O,2023ApJ...956L..23T} and \citet{Banerjee2024}.} The final SEDs were created for 13 different epochs, including the 5 epochs where VHE photons were detected by LHASSO. The MCMC sampling was performed in between 926 and 1926 s since the burst. We have not selected data from very early epochs of GRB~221009A that are too close to the prompt emission. The same model prescriptions were considered for GRB~221009A like the other GRBs, i.e., evolving the $B$-field by keeping the other parameters fixed for all other epochs. We corrected the VHE data sets with {the} EBL and internal absorptions. We note that our models over-predicts the late-time radio data in most of the time intervals in both the ISM and wind scenarios.

GRB~221009A was the most energetic burst in our sample, $E_{\rm k,iso} = 1.5 \times 10^{55}$ and $1.2 \times 10^{55}$ erg, although $\Gamma_0$ is rather moderate as mentioned in Table \ref{tab:bestfit_params}, using $t_{\rm dec}\equiv T_{p}=230$~s \citep{Frederiks_2023} for the ISM and wind cases. The wind medium is preferred over ISM for GRB~221009A based on the BIC values mentioned in Table \ref{tab:BIC_values}. However, it is challenging to distinguish between the wind and ISM scenarios from the SED fittings and the corner plots shown in Figs. \ref{fig:sed_221009a}, \ref{fig:mcmc_221009a_ism}, \ref{fig:corner_221009a_ism}, \ref{fig:mcmc_221009a_wind} and \ref{fig:corner_221009a_wind}.

\begin{table*}
\centering
\caption{A summary of the model parameter values found that best fit the observations for each combination of GRB and density scenario. Bold values were held fixed during the fitting process, {except for $\Gamma_0$ and $\epsilon_B$ which were calculated based on other parameters (see the main text for details)}. {We assumed $E_{\rm k, iso}$ to be five times the isotropic-equivalent $\gamma$-ray energy release $E_{\gamma,\rm iso}$.}}
\label{tab:bestfit_params}
\resizebox{\linewidth}{!}{%
\begin{tabular}{l@{\hspace{3pt}}c@{\hspace{3pt}}c@{\hspace{6pt}}c@{\hspace{3pt}}c@{\hspace{6pt}}c@{\hspace{3pt}}c@{\hspace{6pt}}c@{\hspace{3pt}}c@{\hspace{6pt}}c@{\hspace{3pt}}c}
\hline
Parameter & \multicolumn{2}{c}{180720B} & \multicolumn{2}{c}{190114C} & \multicolumn{2}{c}{190829A} & \multicolumn{2}{c}{201216C} & \multicolumn{2}{c}{221009A} \\
& ISM & Wind & ISM & Wind & ISM & Wind & ISM & Wind & ISM & Wind \\
\hline
$\boldsymbol{E_{\rm k, iso}}$ (erg) & {$6\!\times\!10^{53}$} & $1\!\times\!10^{54}$ & $8\!\times\!10^{53}$ & $8\!\times\!10^{53}$ & $1\!\times\!10^{52}$ & $1\!\times\!10^{52}$ & {$6\!\times\!10^{53}$} & {$6\!\times\!10^{53}$} & {$1.5\!\times\!10^{55}$} & {$1.2\!\times\!10^{55}$} \\
$\boldsymbol{\Gamma_0}$ & {246} & 129 & 167 & 107 & 88 & 30 & {362} & {86} & {129} & {100} \\
$\boldsymbol{t_{\rm dec}}$ (s) & \multicolumn{2}{c}{78} & \multicolumn{2}{c}{116} & \multicolumn{2}{c}{63} & \multicolumn{2}{c}{60} & \multicolumn{2}{c}{230} \\
$\boldsymbol{\dot{M}_{\rm w}}$ (M$_\odot$/yr) & -- & $1\!\times\!10^{-5}$ & -- & $1\!\times\!10^{-5}$ & -- & $5\!\times\!10^{-5}$ & -- & {$5\!\times\!10^{-5}$} & -- & {$8\!\times\!10^{-5}$} \\
$\boldsymbol{v_{\rm w}}$ (cm/s) & -- & $1\!\times\!10^{8}$ & -- & $1\!\times\!10^{8}$ & -- & $2\!\times\!10^{8}$ & -- & {$1\!\times\!10^{8}$} & -- & {$1\!\times\!10^{8}$} \\
$\boldsymbol{A_*}$ & -- & 1 & -- & 1 & -- & 2.5 & -- & {5} & -- & {8} \\
$\boldsymbol{n_0}$ (cm$^{-3}$) & {$5\!\times\!10^{-1}$} & -- & 5 & -- & 30 & -- & {$1\!\times\!10^{-1}$} & -- & {30} & -- \\
$\boldsymbol{\epsilon_B}$ & {$3.5\!\times\!10^{-3}$} & $2.5\!\times\!10^{-4}$ & $4.2\!\times\!10^{-3}$ & $2.2\!\times\!10^{-3}$ & $2.5\!\times\!10^{-4}$ & $1.3\!\times\!10^{-4}$ & {$6.1\!\times\!10^{-3}$}& {$7.1\!\times\!10^{-3}$} & {$5.9\!\times\!10^{-5}$} & {$5.4\!\times\!10^{-5}$} \\
$\epsilon_e$ & {$0.06_{-0.02}^{+0.04}$} & $0.21_{-0.07}^{+0.07}$ & $0.018_{-0.002}^{+0.002}$ & $0.008_{-0.001}^{+0.002}$ & $0.046_{-0.007}^{+0.011}$ & $0.013_{-0.005}^{+0.005}$ & {$0.05_{-0.01}^{+0.02}$} & {$0.054_{-0.006}^{+0.011}$} & {$0.009_{-0.002}^{+0.004}$} & {$0.009_{-0.002}^{+0.005}$} \\
$E_b$ (TeV) & {$0.046_{-0.016}^{+0.04}$} & $0.35_{-0.15}^{+0.27}$ & $0.003_{-0.001}^{+0.002}$ & $0.0006_{-0.0002}^{+0.0003}$ & $0.086_{-0.013}^{+0.03}$ & $0.037_{-0.009}^{+0.016}$ & {$0.027_{-0.005}^{+0.015}$} & {$0.027_{-0.004}^{+0.016}$} & {$0.04_{-0.01}^{+0.02}$} & {$0.05_{-0.01}^{+0.02}$} \\
$p$ & {$2.0_{-0.5}^{+0.5}$} & $2.4_{-0.5}^{+0.4}$ & $1.60_{-0.04}^{+0.04}$ & $1.57_{-0.09}^{+0.07}$ & $2.16_{-0.05}^{+0.05}$ & $2.09_{-0.11}^{+0.11}$ & {$1.9_{-0.1}^{+0.1}$} & {$1.9_{-0.1}^{+0.1}$} & {$2.07_{-0.08}^{+0.15}$} & {$2.04_{-0.07}^{+0.15}$} \\
$E_c$ (TeV) & {$15_{-11}^{+20}$} & $21_{-7}^{+26}$ & $6_{-3}^{+5}$ & $2_{-1}^{+6}$ & $32_{-12}^{+12}$ & $37_{-09}^{+20}$ & {$16_{-2}^{+4}$} & {$17_{-2}^{+3}$} & {$15.2_{-1.7}^{+3}$} & {$19_{-5}^{+9}$} \\
$B$ (G) & {$0.10_{-0.02}^{+0.02}$} & $0.027_{-0.007}^{+0.009}$ & $2.7_{-0.6}^{+0.5}$ & $5.7_{-1.0}^{+1.8}$ & $0.155_{-0.02}^{+0.013}$ & $0.21_{-0.04}^{+0.04}$ & {$0.16_{-0.03}^{+0.07}$} & {$0.18_{-0.02}^{+0.02}$} & {$0.42_{-0.03}^{+0.03}$} & {$0.35_{-0.06}^{+0.06}$} \\
\hline
\end{tabular}%
}
\end{table*}

\section{Discussion \& Conclusions} \label{sec5:concl}

Since the first detection of VHE photons from {GRB~180720B} and GRB~190114C, several models have been proposed to date to explain the {emission site(s) and} radiation mechanism conundrum associated with the VHE GRBs. We discussed some key features of these models below.

\begin{itemize}
    \item Firstly, single-zone analytical models were introduced considering the Thompson and Klein-Nishina cross sections by \citet{derishev2021ApJ.135D, Joshi2021}. \citet{2019Natur.575..455M} employed a detailed synchrotron and SSC model for ISM medium with EBL correction and Klein-Nishina cross section to explain the multiwavelength datasets of GRB~190114C.
    \item \citet{Salafia2022} used the multi-zone structured jet model with off-axis scenarios considering both forward and reverse shocks for GRB~190829A. The study by \citet{Abdalla2019} tried to fit the multiwavelength SED using the \textsc{naima} code for two epochs during the VHE detection. They only focused on the ISM medium without the cooling constraint. Apart from that they considered two emission models: 1) synchrotron with cut-off + SSC; and 2) synchrotron without cut-off. 
    \item The structure jet morphology with core + wing configuration and their transition was considered by \citet{Ren2023} for GRB~221009A. VHE emission in this study was explained by the SSC emission from the core of the jet. In contrast, the low-energy emission comes from forward and reverse shocks of the wing component. \citet{Banerjee2024} adopted the leptonic module of LeHaMoC\footnote{https://github.com/mariapetro/LeHaMoC}, called LeMoC which covers time-dependent synchrotron + SSC model with self absorption, adiabatic loss consideration, $\gamma\gamma$ pair production {and} proton-proton inelastic collisions. {For the same GRB, \citet{2025arXiv250219051S} interpreted the multiwavelength afterglow emission using a two-component jet model, comprising two uniform jets with distinct opening angles—one narrow and one wide. They concluded that the narrow jet in GRB 221009A possesses a smaller opening angle, which accounts for its exceptionally large isotropic-equivalent energy.}
\end{itemize}

Aforementioned studies of VHE GRBs have mostly focused on individual events, each analysed with distinct modelling approaches. As a result, direct comparisons {among model parameters} and population insights have remained pending. Throughout our work, we adopt a unified modelling framework applied consistently to {the current} sample of VHE detected long-durations. This approach not only ensures a methodological coherence but also enables a population study.
Our aim is to study the correlations among the model parameters of VHE GRBs and compare them with the larger sample of long-duration GRBs, offering new insights into their physical properties and emission mechanisms. In our study, we performed a comprehensive multi-epoch panchromatic SED modelling of VHE GRBs using the \textsc{naima} {code for a one-zone emission site}. \textsc{naima} incorporates the simple approach with the combination of synchrotron and IC emission in the forward shock regime for both the ISM and wind environments. It is a time-independent 
code that deals with the full Klein-Nishina cross section and the Bayesian-based MCMC optimisation, which is explained in detail in Appendices \ref{sec:appA_naima} and \ref{sec:appB_MCMC}. We also implemented the EBL correction following \citet{dominguez_extragalactic_2011} in the \textsc{naima} {code}.

In this work, we use multiwavelength datasets of VHE GRBs which are provided in SEDs in the literature. Combining the model with the multiwavelength datasets, spanning from radio to VHE $\gamma$-rays, we constrained five model parameters (the fraction of shock energy transferred to the particles $\epsilon_e$, break energy in electron distribution $E_b$, $E_c$, electron spectral index $p$, and {shock} $B$-field) via MCMC optimisation. Table \ref{tab:bestfit_params} shows the best-fit values of these parameters together with parameters ($E_{\rm k,iso}$, $t_{\rm dec}$, mass-loss rate ${\dot M}_w$, wind speed $v_w$, $A_*$ and $n_0$) that were kept fixed or derived ($\Gamma_0$ and $\epsilon_B$) from other parameters. Three GRBs, i.e., GRB~180720B, GRB~190829A, and GRB~201216C, in our sample favour an ISM environment, while the two bright GRBs, i.e., GRB~190114C and GRB~221009A, favour a wind-like profile. Among the VHE-detected long-durations, GRB~180720B, GRB~190114C, and GRB~201216C have comparable isotropic equivalent energies ($\sim 10^{54}$ erg). In contrast, GRB 221009A stands out as the most energetic burst in the sample, while GRB 190829A represents the least energetic event. One common trend found among all the VHE detected GRBs is the low $B$-field strength and $\epsilon_B$, although with values within the typical range ($ \sim 10^{-6}$ - $10^{-2}$) reported in afterglow studies of other long-duration GRBs \citep{2014Santana, 2016Laskar, Aksulu2022}. GRB~221009A exhibits the lowest $\epsilon_B$ value for the wind medium suggesting a highly inefficient $B$-field amplification in the shocked circumburst medium. Such low $\epsilon_B$ values enhance the SSC contribution, which in turn provide a natural explanation for the observed VHE emission \citep{wang2010ssc, fraija2019ssc}. The parameter $\Gamma_0$ varies widely from $\sim 30$ to 370 without exhibiting any trend, however, the ISM {medium in general requires} higher $\Gamma_0$ values compared to the wind {medium}. VHE GRBs except for GRB~180720B and 201216C, occurred in denser ambient medium. The inferred mass-loss rates from the progenitor stars, in the wind scenario, are consistent across the VHE GRB sample, with GRB~221009A exhibiting the highest value, indicative of a dense stellar wind environment.

\begin{figure}
    \centering
	\includegraphics[width=\columnwidth]{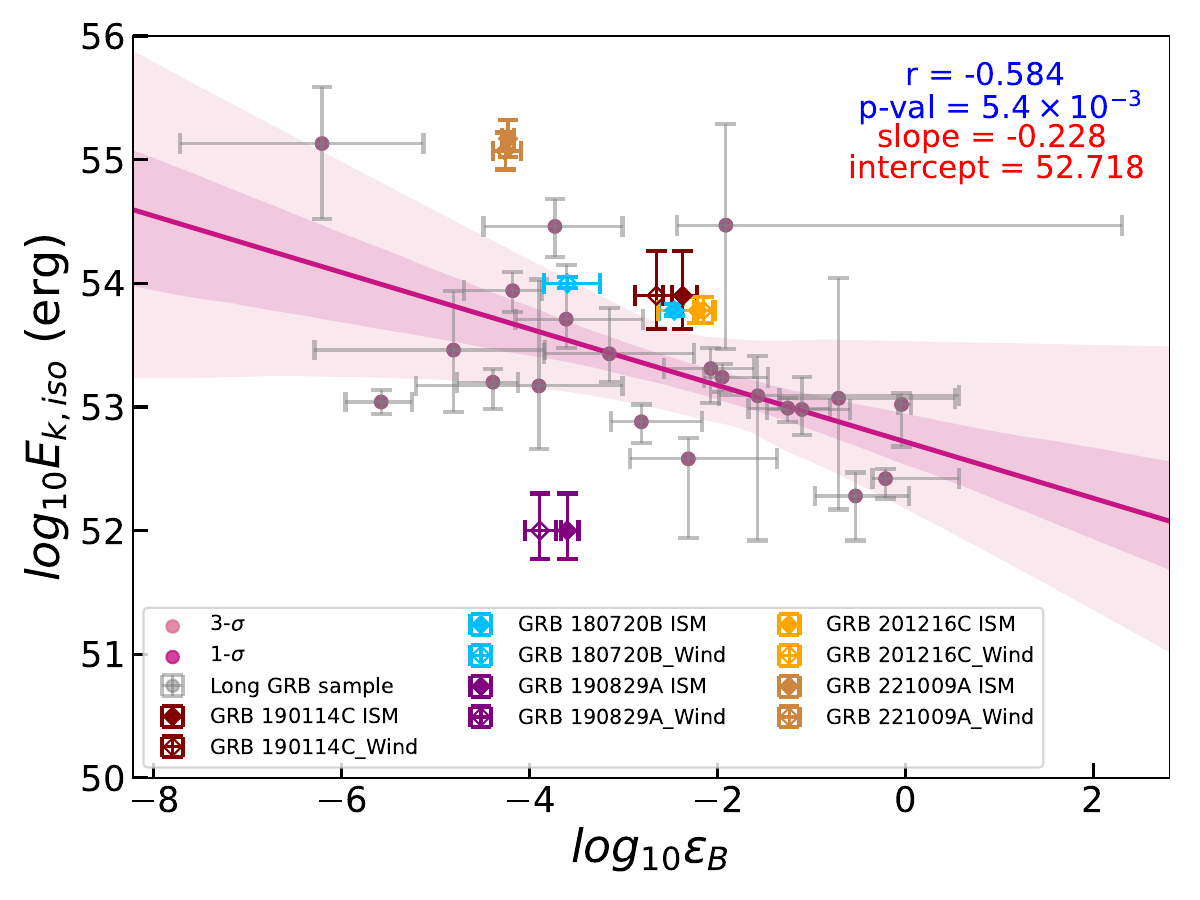}
    \caption{Correlation plot between $\epsilon_B$ and $E_{\rm k,iso}$, for the high energetic GRB sample. The diamonds and circles represent VHE detected long-durations and other high energetic long-durations mentioned in \citet{Cenko2011, Aksulu2022}. The shaded regions represent {$3\sigma$ and $1\sigma$ confidence interval respectively.}}
    \label{fig:corr}
\end{figure}

Finally, we have investigated the correlation between the $E_{ \mathrm{k,iso}}$ and $\epsilon_B$ across a sample of VHE GRBs and {compare with} other high-energy GRBs. This high-energy GRB sample was acquired from \citet{Cenko2011, Aksulu2022}. Our analysis reveals a statistically significant anti-correlation between these two parameters, with a Pearson correlation coefficient $r = -0.584$ and a $p$-value of $5.4 \times 10^{-3}$. This trend is depicted in Fig.~\ref{fig:corr}, where the shaded regions represent the $1\sigma$ and $3\sigma$ confidence interval around the trend-line represented as 

\begin{eqnarray} \label{eqcorr}
    \log_{10} (E_{k, \mathrm{iso}}/{\rm erg}) = -0.228\, \log_{10} (\epsilon_B) + 52.718\,.
\end{eqnarray}

Although most VHE GRBs are intrinsically very energetic and lie above the trend-line, they do not follow the trend of {other} energetic GRBs, {falling outside the $3\sigma$ confidence region - except for the wind model values of GRB~180720B, which is at the edge}. This deviation suggests that VHE GRBs may have a distinct origin or underlying physical mechanism. An exception is GRB~190829A, a lower-luminosity GRB, which, as expected, does not fall within this plane, underscoring its unique position among the VHE GRB population.

In conclusion, we find that a single-zone forward-shock model can adequately fit optical to VHE $\gamma$-ray data for all the VHE GRBs in most of the epochs. {Our modelling over-predicts the radio flux in the earlier epochs of GRB~190829A, and the later epochs of GRB~201216C and GRB~221009A, when data are available.} For GRB~190114C, our modelling can adequately fit optical to VHE $\gamma$-ray data in all epochs, except for the \textit{Fermi}-LAT data during the earliest epoch. Deviation from a single-zone synchro-Compton scenario that we have explored can include 1) an additional emission component; 2) a more complex jet structure; and 3) a more complex density profile of the surrounding environment; or a combination of these. A larger sample of VHE GRBs in the CTA-era \citep{CTAConsortium:2017dvg} will be crucial in modelling VHE GRBs and to learn in more details the nature of these energetic events.

\section*{Acknowledgements}
{We thank Marc Klinger and Lauren Rhodes for useful comments.} This research was supported by the National Research Foundation (NRF) of South Africa through a BRICS Multilateral Grant with number 150504. This work was also supported by the NRF of South Africa through a Postdoctoral Fellowship Grant with number 2205056709.

\section*{Data Availability}

All data used in this work are publicly available from other papers and the \textit{Swift} website \url{https://www.swift.ac.uk/}.



\bibliographystyle{mnras}
\bibliography{vherefs} 




\appendix

\section{Naima afterglow emission model} \label{sec:appA_naima}

We adopted a modified version of the GRB emission code used to analyse the multiwavelength afterglow emission of the H.E.S.S. detection of GRB~190829A \citep{HESS:2021dbz}. We applied this GRB code to fit multiwavelength data at different epochs from several GRBs detected in the VHE range at late times (Section~\ref{sec1:intro}), to understand the underlying particle acceleration and the blastwave evolution processes with time and distance. 

This code utilises \textsc{naima} software as a tool to compute the non-thermal emission, i.e., synchrotron, and IC from a homogeneous distribution of relativistic particle populations of electrons and protons, assuming an arbitrary user-defined particle distribution function. These radiation models are subsequently fitted to the observed multiwavelength spectra using MCMC optimisation, yielding probability distribution functions for the best-fit particle parameters (for more details see Appendix~\ref{sec:appB_MCMC}).

\subsection{GRB blastwave evolution} \label{sec:blastwave}

This section outlines the fundamental expressions for GRB afterglows and the physical parameters used in the code to model their emission, focusing on the underlying assumptions and calculation methods related to the blastwave environment.

Consider a blastwave propagating into the surrounding medium with an isotropic-equivalent energy $E_{\rm k,iso}$ (in erg), a surrounding material density $n$, and observed at a time $t$ after the GRB trigger. Two environmental scenarios are considered: (i) a constant-density interstellar medium (ISM), and (ii) a stellar wind environment in which the density follows an $R^{-2}$ dependence on the radius $R$ of the emitting shell. A third possible scenario is an average between the first two scenarios, although not used in this study. Given these initial physical parameters of the blastwave, the Lorentz factor $\Gamma$ of the forward shock is computed using the formalism of \citet{Blandford1976}, in which $\Gamma^2=E_{\rm iso}/Mc^2$, where $M$ is the mass of the material swept up by the shock. Depending on the scenario chosen, $\Gamma$ is determined as follows, 

\begin{eqnarray} \label{eq:lorentz_factor}
\Gamma(t) =
 \begin{cases}
 \left(\frac{1}{8}\right)^{3/8}\left(\frac{3E_{\rm iso}}{4\pi n(R) m_p c^2 (ct)^3}\right)^{1/8} \,; & {\rm ISM} \\
 \left(\frac{3E_{\rm iso} v_w}{4c^3 t \dot{M}_w}\right)^{1/4} \,; & {\rm wind\,,}
\end{cases}
\end{eqnarray}

where $\dot{M}_{w}$ (in solar mass $M_{\odot}$ per yr) is the mass-loss rate of the progenitor star in a stellar wind with speed $v_{\mathrm w}$ (in km/s). Based on the blastwave dynamics described in \citet{Blandford1976}, $R$ is determined by the relation $R = A_0\Gamma^2 ct$, where $A_0$ is a scenario-dependent constant. Thus, 

\begin{eqnarray} \label{eq:radius}
R(t) = \Gamma^2 c\Delta t \times
 \begin{cases}
 8 \,; & {\rm ISM} \\
 4 \,; & {\rm wind}.
 \end{cases}
\end{eqnarray}

The densities in the two cases can be expressed as functions of $R$ as follow,

\begin{eqnarray} \label{eq:density}
n(R) = 
 \begin{cases}  
 n_0 \,; & {\rm ISM} \\
 \frac{\dot{M}_w}{4\pi v_{\rm w}R^2m_p} \,; & {\rm wind}.
 \end{cases}
\end{eqnarray}

\subsection{Particle distribution and acceleration} 
\label{sec:particle_dist}

The electron distribution is specified prior to computing the particle emission in the blastwave and is assumed to follow an exponential cutoff broken power law (ECBPL) as implemented in \textsc{naima}, and is expressed as follows

\begin{eqnarray} \label{eq:ECBPL}
f(E) = e^{-(E / E_c)^\beta}
    \begin{cases}
    A_0 (E / E_0) ^ {-\alpha_1};    & E < E_b \\
    A_0 (E_b/E_0) ^ {\alpha_2-\alpha_1} (E / E_0) ^ {-\alpha_2}; & E > E_b. \\
    \end{cases}
\end{eqnarray}

Here $E_c$ is the cutoff energy, $E_0$ is the reference energy (chosen as 1 TeV), $E_b$ the break energy, $\alpha_1$ and $\alpha_2$ are the spectral indices below and above $E_b$, and $A_0$ is the normalisation. All these quantities are defined in the shock frame.

To model the electron distribution in \textsc{naima} some physical constraints need to be considered. These include:
\begin{itemize}
    \item \textit{Calculate of the minimum injection energy $E_{\rm min}$:} this minimum energy is not fixed a priori but depends on the fraction of shock energy transferred to the particles, denoted by $\eta_e$ (equivalent to $\epsilon_e$). Using this parameter, $E_{\rm min}$ can be derived from energy and number conservation integrals over the particle distribution. Assuming the injection spectral index $p_{\rm inj}>2$ and $E_{\rm min}\ll E_{\rm max}$, the resulting expression relates $E_{\rm min}$ to $\eta_e\Gamma m_p c^2$. To ensure numerical stability during MCMC sampling, the ratio of energy to particle number is computed numerically using the full electron distribution. An iterative method adjusts $E_{\rm min}$, starting from 1~GeV, to bring the ratio close to unity. Typically, a single iteration is sufficient for convergence within a factor of 2, balancing accuracy and computational efficiency. 
    \item \textit{Normalisation of the electron distribution:} the free parameter $\eta_e$ relates to the actual normalisation of the distribution. The process begins with calculating the internal energy density of the shocked plasma, given by $w=2\Gamma^2n_0m_pc^2$. Based on the values of $\alpha_1$ and $\eta_e$, the value of $E_{\rm min}$ is determined, and a unit-normalised electron energy distribution is constructed. The total energy density of this unit distribution is then calculated using the integral $ T=1/V\int_{E_{\rm min}}^{E_{\rm max}} E dN/dEdE$, where the volume $V$ is that of a spherical shell $V=4\pi R^2(R/(9 \Gamma))$ with the numerical factor indicating the width of the shell depending on the environment, e.g., $R/9\Gamma$ (for ISM) or $R/3\Gamma$ (for wind). The actual normalisation factor $A_0$ is then found by scaling the unit-normalised distribution so that its energy density matches the physical energy content, given by $A_0=\eta_e w/T$. Once $A$ is determined, the electron distribution is fully specified and can be used by \textsc{naima} to compute the resulting radiation. 
    \item \textit{Age of the system:} the age of the system is given by the Lorentz-factor corrected time from trigger, $t^\prime_{\rm cool} = (t-t_0)\Gamma$ where the time has been corrected for the relativistic boost (primed quantities are in the shock frame). This means that the cooling break should be in a position in the spectrum for which the energy of the particles have the same cooling time $t_{\rm cool}$ as the age of the system. For a generic GRB, using these relations can give some constrains on the intensity of the $B$-field imposing that the cooling time of the electrons at the break is at the same level of the age of the system. This constraint is incorporated into the prior function and can be enabled during initialization by setting the cooling constraint option to \textit{true} (see Appendix~\ref{sec:appB_MCMC}).
    \item \textit{Calculate the internal absorption:} the code takes into account the $\gamma\gamma$-absorption that affects the emitted high-energy photons when they interact with the synchrotron photons inside the source (internal absorption). To take this effect into account, we need the number density of the target radiation field and the cross section of the absorption process. The number density of the target radiation field $n_{\rm ph}$ is actually already computed by \textsc{naima} when determining the SSC component. The cross section for the process is the analytical approximation coded in \citet{Aharonian2004}. In this way the optical depth parameter $\tau$ is computed (see Section~\ref{sec:gamma_gamma_abs}). Assuming that the size of the region in which absorption and emission happens is the width of the shell, depending on the scenario. In the code there are two different implementations for the absorption. The default one takes into account that in the same region we have both emission and absorption, i.e., $F = (F^{\prime}/\tau) (1-e^{-\tau})$ where $F^{\prime}$ indicates the intrinsic flux and $F$ the observed flux \citep{1979rpa..book.....R}. Alternatively, there is also the implementation of $F = F^{\prime} e^{-\tau}$. In this study we used the latter one. 
\end{itemize}

Once the electron distribution is determined, the population of electrons in the shocked gas is assumed to be accelerated in the $B$-field in the shocked region. The ratio between a given $B$-field energy and shock energy is given as $\epsilon_B$, and is expressed as a function of a given $B$-field strength as follows

\begin{equation} \label{eq:epsB}
\epsilon_B = \frac{B^2}{16\pi\Gamma^2 n m_p c^2},
\end{equation} 

with initial parameters including a temporary amplitude ($1$ eV$^{-1}$), initial reference energy ($E_0$ in TeV), $E_b$, $\alpha_1$, $\alpha_2$, $E_c$, and the exponential cutoff rapidity $\beta=1$. Once the ECBPL parameters are defined for a certain energy range with the maximum energy equal to $E_c$, and after the calculation of the $E_{\rm min}$ as a function of $\eta_e$ and $\Gamma$, the synchrotron emission can be determined (see Section~\ref{sec:synchrotron}). Using this calculated synchrotron emission, a new amplitude as a function of $\eta_e$, $V$ where the emission takes place, and shock energy can be calculated and used to compute the total energy in the electron distribution.

\subsection{Synchrotron radiation} \label{sec:synchrotron}

\textsc{naima} accounts for the maximum energy of electrons for synchrotron emission by using the effect of synchrotron burn-off, which is due to the balancing of acceleration and synchrotron losses of electrons in the $B$-field (in units of Gauss), given by Eq.~[18] from \citet{Aharonian2000} as

\begin{eqnarray} \label{eq:Eburnoff_lim}
    E_{c,\rm limit} & = \left(\frac{3}{2}\right)^{3/4}\sqrt{\frac{1}{e^3 B}} m_{\rm e}^2 c^4 \eta^{-1/2} \,;~~~~ {\rm TeV}
\end{eqnarray}
with all parameters in the cgs units. Here the acceleration efficiency parameter $\eta$ is assumed to be $\geq1$. 

The synchrotron cooling time of the electrons with energy $E_b$ (in eV) produced in the $B$-field follows from Eq~[1] in \citet{Aharonian2000}:

\begin{eqnarray} \label{eq:tcool}
    t_{\rm cool} & = \frac{6 \pi m_{\rm e}^4 c^3}{\sigma_{\rm T} m_{\rm e}^2 E_b B^2 (1+Y)} \,;~~~~ {\rm s}
\end{eqnarray}

with $\sigma_{\rm T}$ the Thomson cross-section in cm and the dimensionless Compton parameter $Y$. The cutoff characteristic energy of a synchrotron photon at the electron cutoff energy and a specific $B$-field strength (see Eq~[3.30] from \citet{Aharonian2004} adapted for electrons) is given as

\begin{eqnarray}\label{eq:charE}
    h\nu_{c} = \sqrt{\frac{3}{2\myphantom}} \frac{heBE_c^2}{{2\myphantom\pi m_{\rm e}^3 c^5 (1+Y)}} \,;~~~~ {\rm eV}
\end{eqnarray}

with $h$ the Planck constant, and $\nu_{\rm c}=(3/2)\nu_{\rm L}(E/m_{\rm e}c^2)^2$ the characteristic frequency of the synchrotron radiation emitted by an electron, with $\nu_{\rm L}=eB/2\pi m_{\rm e}c$ the Larmor frequency. This cutoff photon energy is then used to determine the synchrotron luminosity $L_{\rm sy}$ (in eV/s). \textsc{naima} uses a thin shell, spherical approximation to calculate the number density of synchrotron photons in the considered emission region. Therefore, no correction factor is needed \citep[see, e.g.,][]{Atoyan1996} and thus

\begin{eqnarray} \label{eq:photon_density}
n_{\rm sy} = \frac{L_{\rm sy}}{4\pi R^2 c}
\end{eqnarray}
in eV$^{-1}$ cm$^{-3}$.

For the assumed electron distribution at a given photon energy, and luminosity distance $D_{\rm L}$ from the source, the spectral component is expressed as follows

\begin{eqnarray} \label{eq:sync_sed}
\left(\frac{dN}{dE}\right)_{\rm sy} = \Gamma^2\left(\frac{dN}{dE}\right)_{\rm sy}= \Gamma^2\frac{L_{\rm sy}}{4\pi D_{\rm L}^2}
\end{eqnarray}

where $\Gamma$ is equivalent to the Doppler boosting, and the flux is transformed to the blastwave frame.

\subsection{Synchrotron self-Compton radiation} \label{sec:ssc}

\textsc{naima} assumes a one-zone model to calculate SSC emission, where the target photon field consists of synchrotron photons with a certain energy range and density, as described in Section~\ref{sec:synchrotron}. The synchrotron photon luminosity $L_{\rm sy}$ is computed over a specified energy range and binning. From this, the synchrotron photon number density $n_{\rm sy}$ is obtained using Eq.~(\ref{eq:photon_density}).

For the assumed electron distribution at a given photon energy, and $D_{\rm L}$ from the source, with synchrotron target photon field, the spectral component is expressed as follows,

\begin{eqnarray} \label{eq:ssc_sed}
\left(\frac{dN}{dE}\right)_{\rm SSC} = \Gamma^2\left(\frac{dN}{dE}\right)_{\rm SSC}= \Gamma^2\frac{L_{\rm SSC}}{4\pi D_{\rm L}^2} \,.
\end{eqnarray}

\subsection{External Compton radiation}
\label{sec:ec}

We implemented the IC for external photon fields including the CMB and the far-IR. The seed photon field parameters needed are the photon field considered with the associated temperature and energy density. The EC component is computed in the jet frame for each field and then added together to get the total EC flux. The temperature for the FIR is taken as $30\,\Gamma$ (in K), witha an energy density of $0.5\,\Gamma^2$ in (eV / cm$^2$). For the CMB field, we assumed a temperature of $2.72\,\Gamma$ (in K), and its energy density of $0.260\,\Gamma^2$ in (eV~ cm$^{-2}$). The total EC flux is included in the model with the other components previously mentioned.

For the assumed electron distribution at a given photon energy, and $D_{\rm L}$ from the source, the spectral component is expressed as follows

\begin{eqnarray} \label{eq:ec_sed}
\left(\frac{dN}{dE}\right)_{\rm EC} = \Gamma^2\left(\frac{dN}{dE}\right)_{\rm EC}= \Gamma^2\frac{L_{\rm EC}}{4\pi D_{\rm L}^2} \,.
\end{eqnarray}

\subsection{Modified electron cutoff energy}
\label{sec:modify}

We have modified \textsc{naima} for the maximum energy for electrons by including losses by synchrotron, SSC, and EC simultaneously.
Using the expression for the maximum Lorentz factor of the electron distribution $\gamma_s(t)$ in Eq.~[9] from \citet{Barnard2024} and the relation $\gamma=(E_c/m_e c^2)+1$, we could get an estimate for the maximum energy of the electron distribution $E_c$ (in the blastwave frame) using the following expression:

\begin{eqnarray} \label{eq:Ec_est}
E_c = m_e c^2\left[\left(\frac{6\pi e}{\phi \sigma_{\rm T}B(t)(1+Y+\Gamma^2 u_{\rm ext}/u_{\rm B})}\right)^{1/2}-1\right]
\end{eqnarray}

We used $\phi=10$ and $Y\approx\sqrt{\epsilon_e/\epsilon_B}$ in the slow-cooling regime. This estimate constrained the prior value range for the $E_c$ discussed in Sec.~\ref{sec:synchrotron}.

\subsection{Optical depth for \texorpdfstring{$\gamma\gamma$}{gamma-gamma}-absorption} \label{sec:gamma_gamma_abs}

\textsc{naima} uses the $\gamma\gamma$ cross section averaged over scattering angle, with $\epsilon_\gamma$ and $\epsilon_0$ the $\gamma$-ray photon and target photon energy respectively in eV units (from Eq.~[5] in \citet{Eungwanichayapant2009} and \citet{Aharonian2004}), as given by 

\begin{align} \label{eq:cross_section}
\sigma_{\gamma\gamma}
&=\frac{3\sigma_{\rm T}}{2s_0^2} \biggl[\left(s_0 + \tfrac{1}{2}\ln{s_0} - \tfrac{1}{6} + \tfrac{1}{2 s_0}\right)\ln\!\left(\sqrt{s_0} + \sqrt{s_0 - 1}\right) \nonumber \\
&\quad - \left(s_0 + \tfrac{4}{9} - \tfrac{1}{9s_0}\right)\sqrt{1 - \tfrac{1}{s_0}} \biggr] \,[{\rm cm}^2]
\end{align}

where $s_0=\epsilon_\gamma \epsilon_0$, with $s_0 > 1$ is a mask condition to account for the threshold effect, and the cross section is an approximation good within 3\%.

The absorption coefficient $K$ that will then be spatially integrated is $K(\epsilon_\gamma) = \int_{\epsilon_0} \sigma_{\gamma\gamma}(\epsilon_\gamma,\epsilon_0) (dn/d\epsilon_0)d\epsilon_0$, where $dn/d\epsilon_0$ is the spectral number distribution of the target photon field (the inner integral of Eq.~[3.24] of \citet{Aharonian2004} in units of eV$^{-1}$ cm$^{-3}$). The absorption coefficient as (in units cm$^{-1}$) is

\begin{eqnarray} \label{eq:abs_coeff}
K(\epsilon_\gamma) = \int_{\epsilon_1}^{\epsilon_2} \sigma_{\gamma\gamma}(\epsilon_\gamma,\epsilon_0)n(\epsilon_0) \,.
\end{eqnarray}

The corresponding optical depth for $\gamma$-ray absorption, with the assumption of a homogeneous radiation photon field in a source of size $R$ (in cm; from Eq.~[3.24] of \citet{Aharonian2004}), is as follows:

\begin{equation} \label{eq:tau_gamgam}
\tau(\epsilon_\gamma) = \int_{0}^{R} \int_{\epsilon_1}^{\epsilon_2} 
\sigma_{\gamma\gamma}(\epsilon_\gamma,\epsilon_0)\,n(\epsilon_0,r)\,d\epsilon\,dr 
= RK(\epsilon_\gamma,R)
\end{equation}

with an additional spatial dependence to Eq.~(\ref{eq:abs_coeff}), $n(\epsilon,R)$ that describes both the spectral and spatial distribution of the target photon field in the source. \textsc{naima} computes the optical depth in a shell of width $R/(9\Gamma)$ after transformation of the $\gamma$ ray energy of the data in the blastwave frame. The energy and density of the target photon fields are chosen to calculate both for the SSC and EC emission.

The SSC emission is corrected for $\gamma\gamma$ absorption and EBL attenuation using the flux component in Eq.~(\ref{eq:ssc_sed}) and the opacity in Eq.~(\ref{eq:tau_gamgam}) and section~\ref{sec2:model} as follows,  

\begin{eqnarray} \label{eq:ssc_gamgam}
\left(\frac{dN}{dE}\right)_{\rm SSC} = \left(\frac{dN}{dE}\right)_{\rm SSC}\exp^{-\tau_{\gamma\gamma}(\epsilon_\gamma)}\exp^{-\tau_{\rm EBL}(\epsilon_\gamma)}
\end{eqnarray}

The EC emission is corrected similarly as the SSC by using the flux component in Eq.~(\ref{eq:ec_sed}) and the opacity in Eq.~(\ref{eq:tau_gamgam}) and Section~\ref{sec2:model}. The total model after absorption is the summation of the distinct components after corrections.

\section{Optimal parameter space} \label{sec:appB_MCMC}

We obtained the optimal parameter space in which the emission models describe the observations the best by using a MCMC fitting routine including a BIC employing the Python package emcee as the sampler \citep{Foreman2016}.

\subsection{Initialisation of the model parameters and priors} \label{sec:initparams}

The functions take automatically the initialisation parameters: $E_{\rm k,iso}$, $n=n_0$ (if ISM case is selected), observational $\gamma$-ray energy $E_{\gamma,\rm iso}$, the duration of the afterglow observation determined by and average using $t_{\rm start}$ and $t_{\rm stop}$, $z$, scenario (ISM or wind), $\dot{M}_{\rm w}$, $v_{\rm w}$, and prior functions to constrain the model, i.e., \textit{cooling constraint} and \textit{synchrotron limit}, see Appendix~\ref{sec:appA_naima} for more details of the model. Other prior model parameters for the model fit include $\eta_{\rm e}$ ($\equiv \epsilon_{\rm e}$), $E_b$, $E_c$, $\alpha_2$, and the $B$-field. The $\log_{10}$ of all prior parameters are taken except $\alpha_2$. Then the emission from particles are calculated. The index $\alpha_1$ ($=\alpha_2 - 1\equiv p$) of the electron distribution is fixed to be a cooling break, and $\alpha_2$ of the electron distribution is free, and the minimum energy and the normalisation of the electron distribution are derived from the parameter $\eta_{\rm e}$.

Two constraints that are important when considering the cooling of the synchrotron photons in the system are, the \textit{cooling constraint} that adds to the prior a constraint for which the cooling time at break is approximately the age of the system, and is by default true. The second constraint \textit{synchrotron limit} determines whether the model is synchrotron or IC dominated, thus when false the model is a standard (one-zone) SSC model and if true it is a synchrotron dominated model. The latter constraint is by default false. It is important to note that if the model is synchrotron dominated the \textit{cooling constraint} has no effect.

The basic prior function used is determined by the above constraints, with some basic parameters of the electron distribution left free. The priors are uniformly distributed and take as input the $\log_{10}$ value of the parameters and a lower and an upper limit. The lower limit of $E_b$ is the minimum injection energy, and for $E_c$ the upper limit is the maximum limit of the cut-off dictated by the synchrotron burn-off limit; see Eq.~(\ref{eq:Eburnoff_lim}). In this function, the synchrotron cooling time of an electron at the break, to be equal to the co-moving age of the system and is calculated as $t_{\rm age}=\Gamma t$. The cooling time at break energy is calculated using Eq.~(\ref{eq:tcool}) at the given $B$-field (G) and $E_b$ (TeV). This prior on the cooling time at break (i.e., approximately the age of the system) is implemented through a normal prior distribution which takes the two aforementioned times as inputs, with parameter $t_{\rm cool}$, and the range determined by $t_{\rm age}$. Lastly, a prior probability (the sum of all the priors including the additional prior) is returned. 

\subsection{Model fitting and selection of the best fit} \label{sec:fitting_selection}

The MCMC fitting routine that is used by \textsc{naima} runs the sample (a.k.a. the full chain of the MCMC), save the complete run, and stores the best fit results in a table. For each MCMC fitting, the number of parallel walkers, burn-in steps, steps after burn-in, and parallel threads can be chosen. The prefit performs a maximum likelihood (ML) fit to the dataset to get a better starting point for the MCMC chain and is by default true. It is important to note that the ML fit might converge in a region not allowed by the parameter space, which is also limited by the lower and upper boundaries set in the prior functions. The fitting routine takes as input the dataset, the prior model parameters (see Section~\ref{sec:initparams}), the model itself, the probability returned by the prior function, the prefit, the number of walkers, burn-in steps, runs, and threads. After each sample, a set of diagnostic plots is delivered that includes the prior distribution for each model parameter, a corner plot, and the SED that illustrates the model fit to the data and the $1\sigma$-confidence band that corresponds to the corner plot. 

The best-fit model to the data selection is done via the Bayesian inference method through which a given model is selected from a set of possible models. For all competing models, the Bayes factor is computed in order to gauge which model provides a better fit to the multiwavelength data. However, computing the Bayes factor is often non-trivial, and a simpler way to obtain an estimate is using the BIC. Note that the BIC is only a valid approximation for the Bayes factor when the number of data points is much larger than the number of parameters. The BIC is a criterion for model selection among a finite set of models, where models with lower BIC are generally preferred.

\begin{table}
\caption{Comparison of two density scenarios for each GRB. We computed the $\Delta$BIC value to differentiate which fit is better between the competing models, for the same GRB. A lower BIC (boldfaced) indicates a better model, and the $\Delta$BIC, calculated by subtracting the best BIC from other models, helps assess the relative quality of those models. The criteria for $\Delta$BIC is as follows based on evidence against the compared model: $<2$ (negligible), 2-6 (positive), 6-10 (strong), and $>10$ (very strong).}
\label{tab:BIC_values}
\resizebox{0.8\linewidth}{!}{%
\footnotesize
\begin{tabular}{lccc}
\hline
Source & \multicolumn{2}{c}{BIC} & $\Delta$BIC \\
 & ISM & Wind &  \\
\hline
GRB~180720B & \textbf{1382641} & 1527916 & 145275 \\
GRB~190114C & 32139 & \textbf{32011} & 128 \\
GRB~190829A & \textbf{301846} & 375769 & 73923 \\
GRB~201216C & \textbf{241042} & 267734 & 26692 \\
GRB~221009A & 778963 & \textbf{774016} & 4947 \\
\hline 
\end{tabular}
}
\end{table}

The SEDs and corner plots for each GRB studied in the paper are included in this section.

\begin{figure*}
    \centering
    \begin{minipage}[t]{0.45\textwidth}
        \centering
        \includegraphics[width=\linewidth]{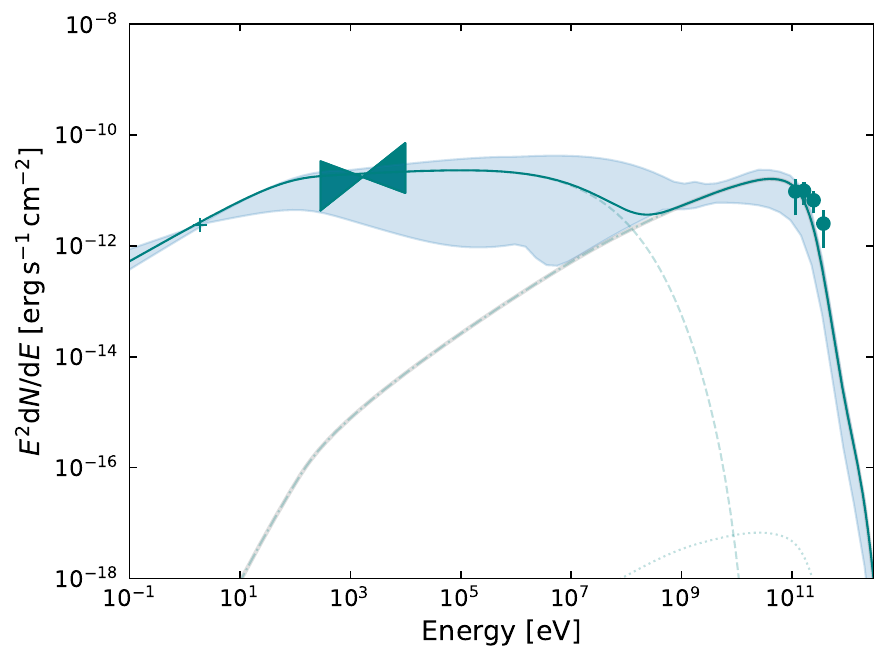}
        \caption{SED fitting of GRB~180720B in the ISM medium. The MCMC $1\sigma$-confidence band is shown over the model for the best-fit parameters are mentioned in Table \ref{tab:bestfit_params}. The synchrotron (dashed), SSC (dash-dotted), EC (dotted), and total (solid) emission are shown. This line style convention is followed consistently for all VHE GRBs under both the ISM and wind-like ambient medium scenarios.}
        \label{fig:mcmc_180720b_ism}
    \end{minipage}
    \hfill
    \begin{minipage}[t]{0.45\textwidth}
        \centering
        \includegraphics[width=\linewidth]{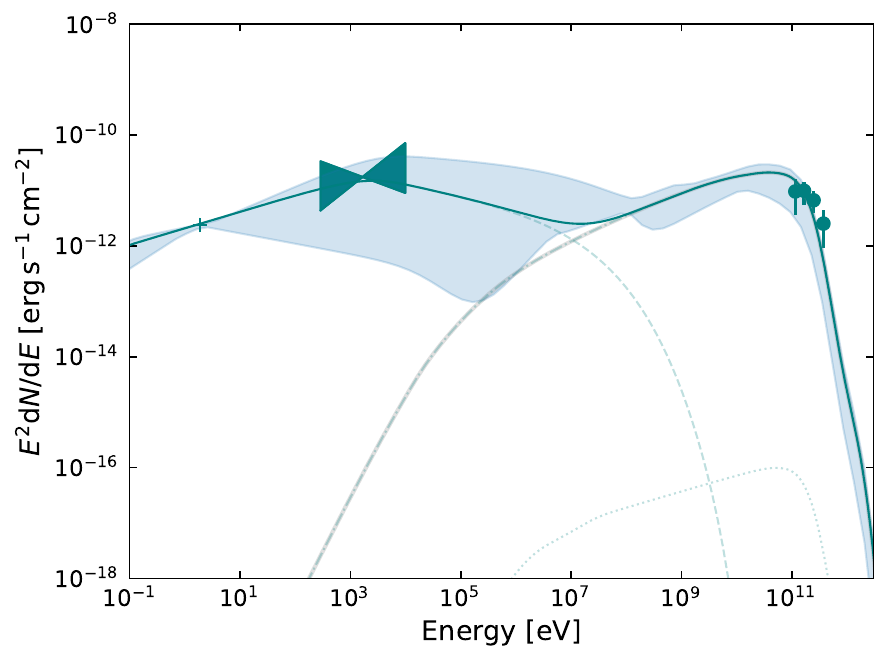}
        \caption{SED fitting of GRB~180720B for the wind medium. The MCMC $1\sigma$-confidence band is shown over the model for the best-fit parameters are tabulated in Table \ref{tab:bestfit_params}}
        \label{fig:mcmc_180720b_wind}
    \end{minipage}
     
\end{figure*}

\begin{figure*}
    \centering
    \begin{minipage}[t]{0.48\textwidth}
        \centering
        \includegraphics[width=\linewidth]{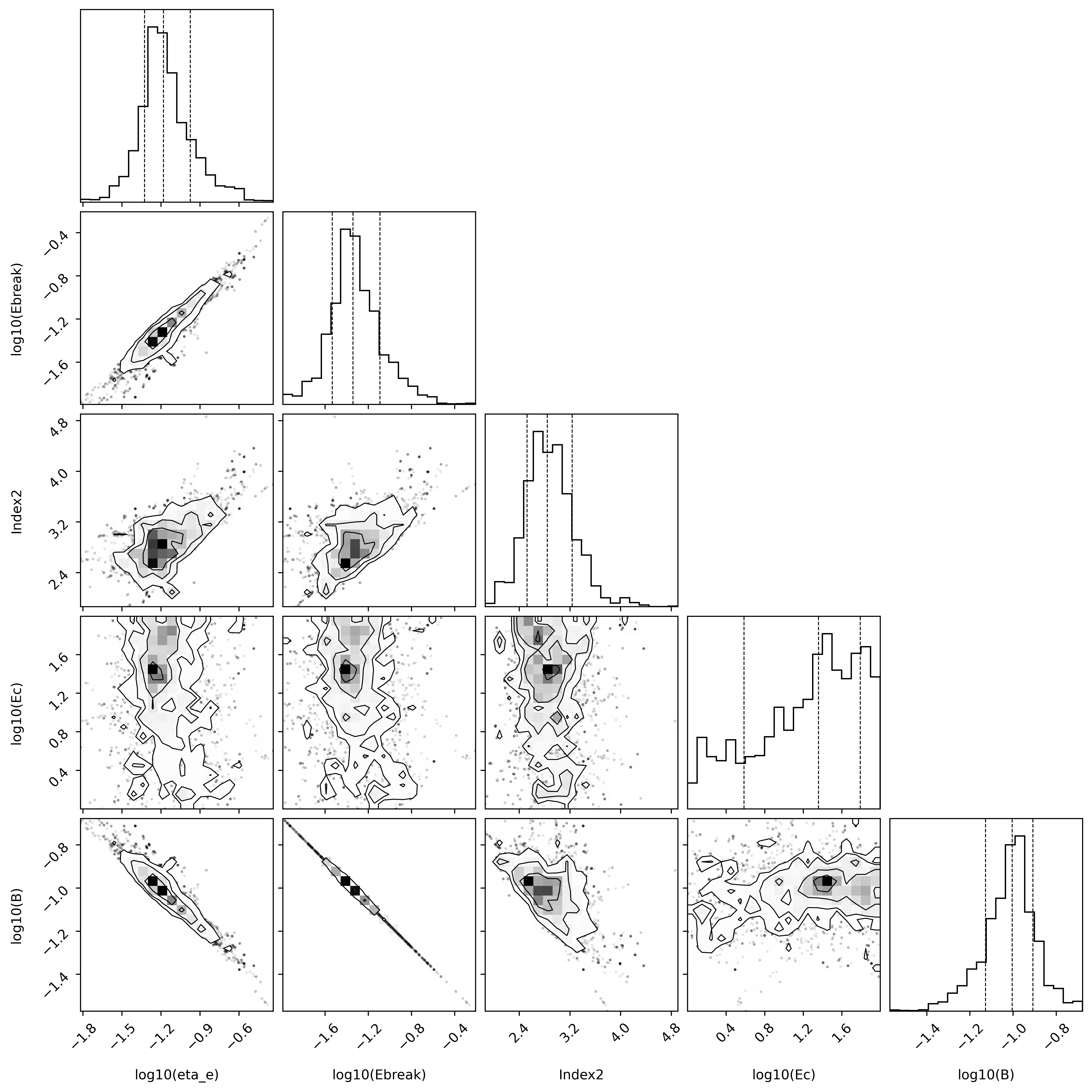}
        \caption{GRB~180720B corner plot for the ISM medium}
        \label{fig:corner_180720b_ism}
    \end{minipage}
    \hspace{0.01\textwidth} 
    \begin{minipage}[t]{0.48\textwidth}
        \centering
        \includegraphics[width=\linewidth]{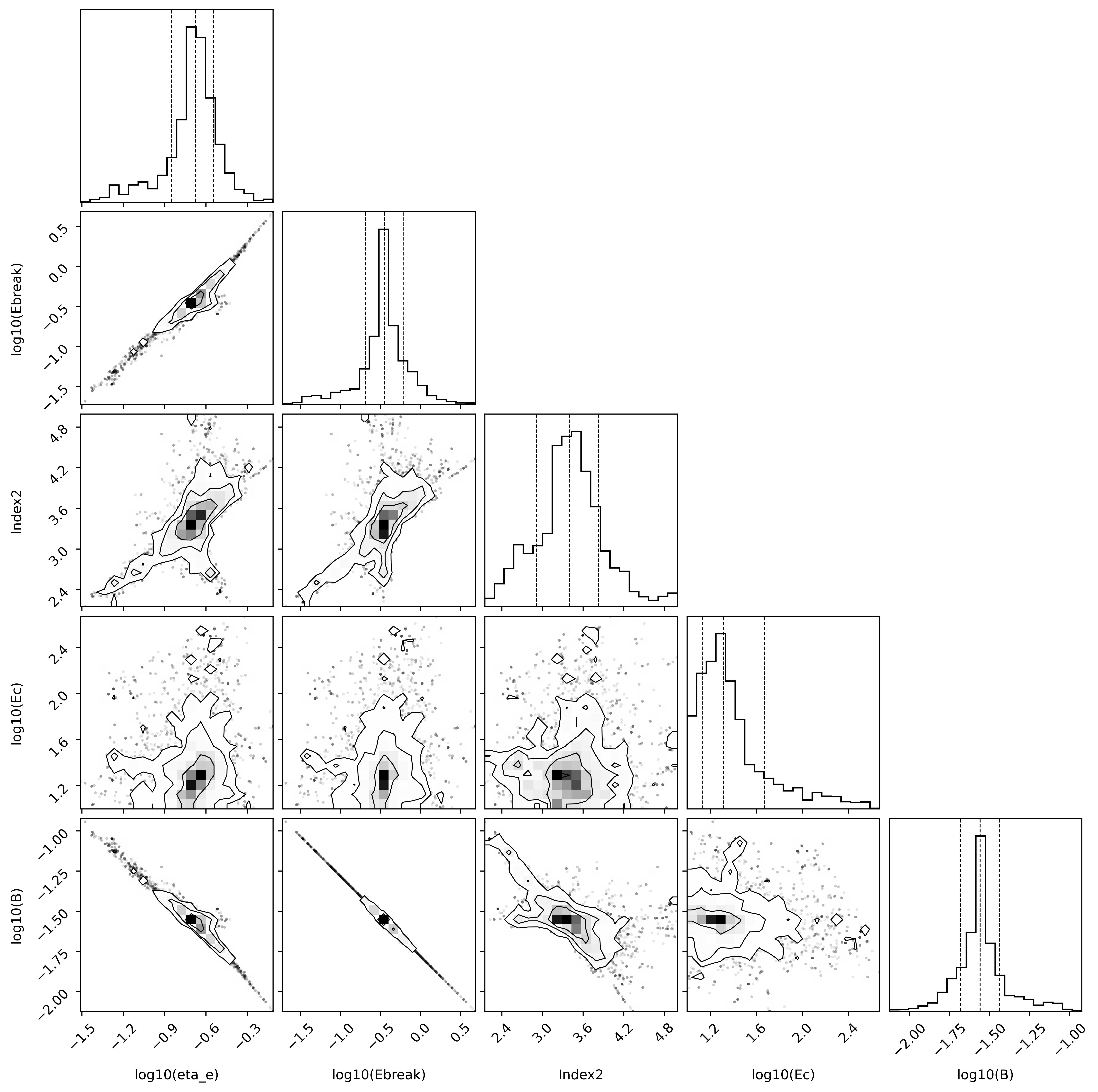}
        \caption{GRB~180720B corner plot for the wind medium}
        \label{fig:corner_180720b_wind}
    \end{minipage}
    
\end{figure*}

\begin{figure*}
    \centering
    \begin{minipage}[t]{0.45\textwidth}
        \centering
        \includegraphics[width=\linewidth]{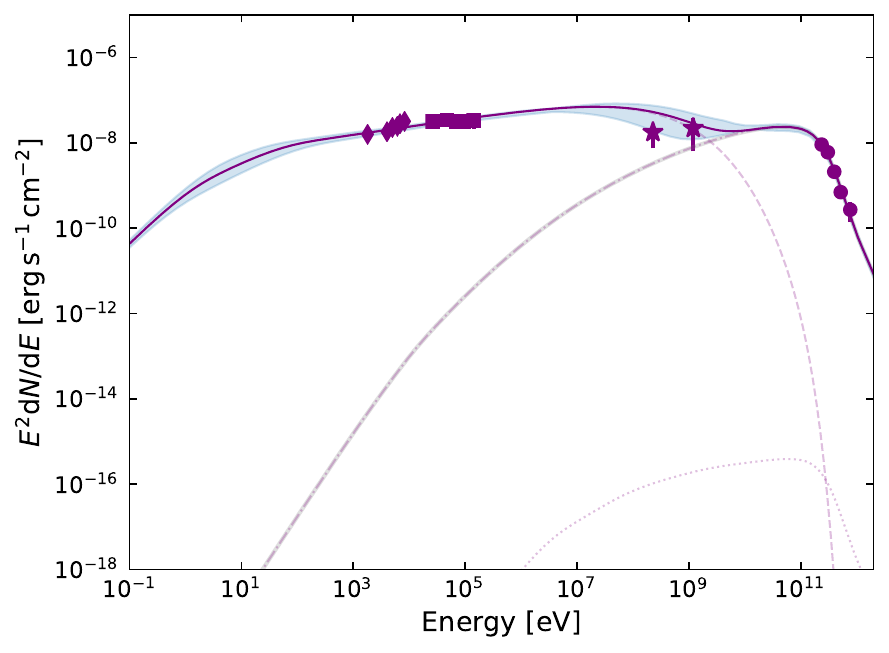}
        \caption{SED fitting of GRB~190114C in the ISM medium. The MCMC $1\sigma$-confidence band is shown over the model for the best-fit parameters are tabulated in Table \ref{tab:bestfit_params}.}
        \label{fig:mcmc_190114c_ism}
    \end{minipage}
    \hfill
    \begin{minipage}[t]{0.45\textwidth}
        \centering
        \includegraphics[width=\linewidth]{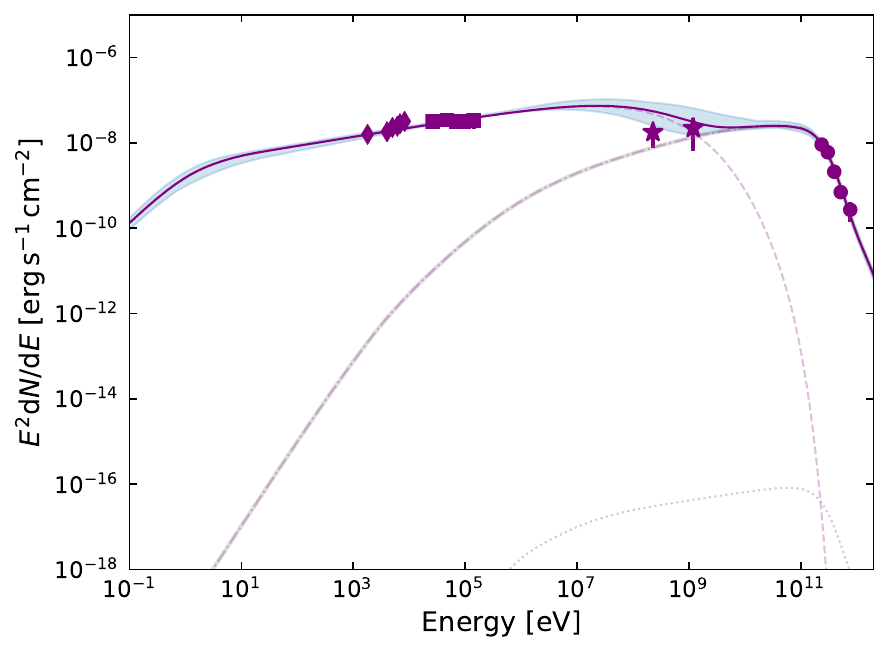}
        \caption{SED fitting of GRB~190114C for the wind medium. The MCMC $1\sigma$-confidence band is shown over the model for the best-fit parameters are tabulated in Table \ref{tab:bestfit_params}.}
        \label{fig:mcmc_190114c_wind}
    \end{minipage}
     
\end{figure*}

\begin{figure*}
    \centering
    \begin{minipage}[t]{0.48\textwidth}
        \centering
        \includegraphics[width=\linewidth]{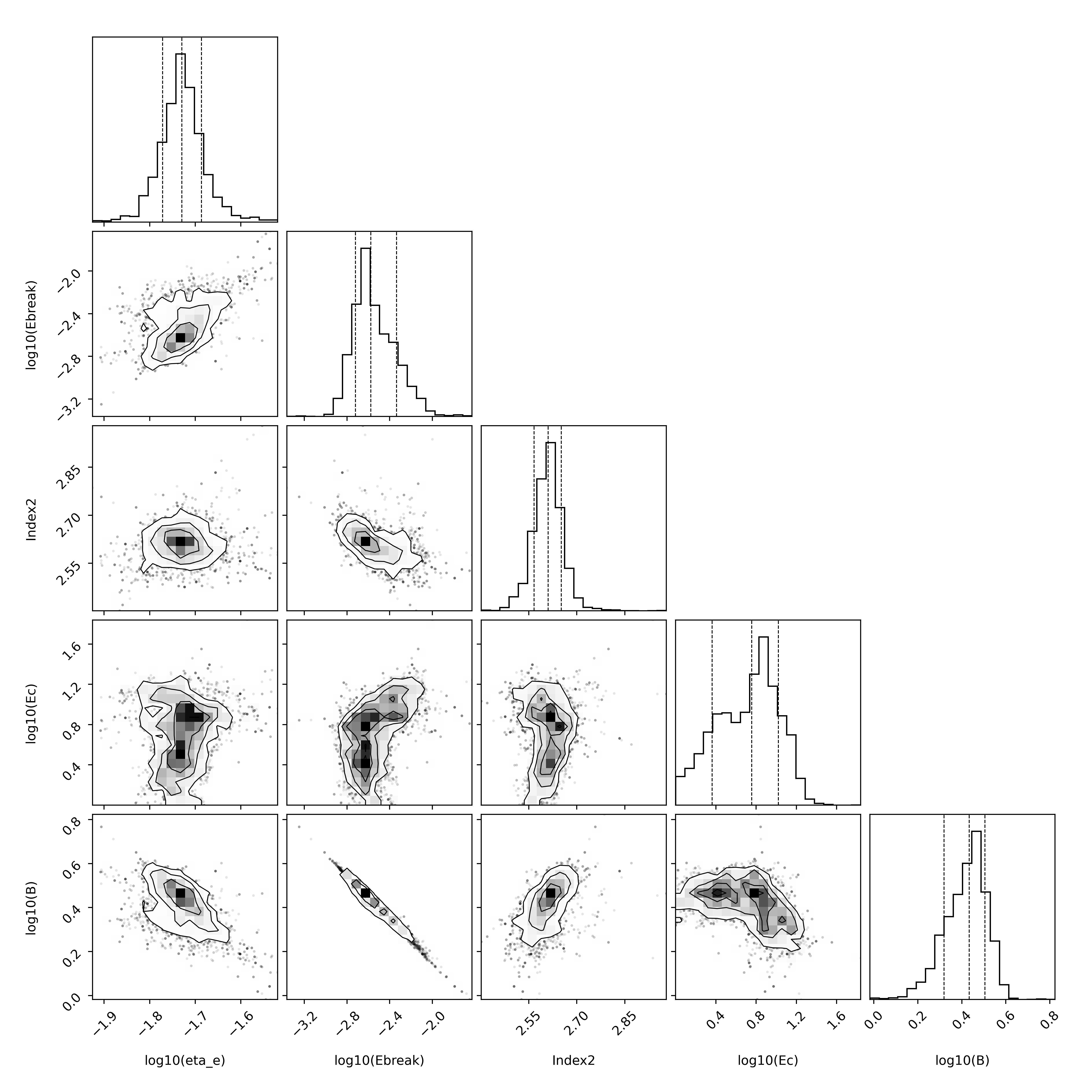}
        \caption{GRB~190114C corner plot for the ISM medium.}
        \label{fig:corner_190114c_ism}
    \end{minipage}
    \hspace{0.01\textwidth} 
    \begin{minipage}[t]{0.48\textwidth}
        \centering
        \includegraphics[width=\linewidth]{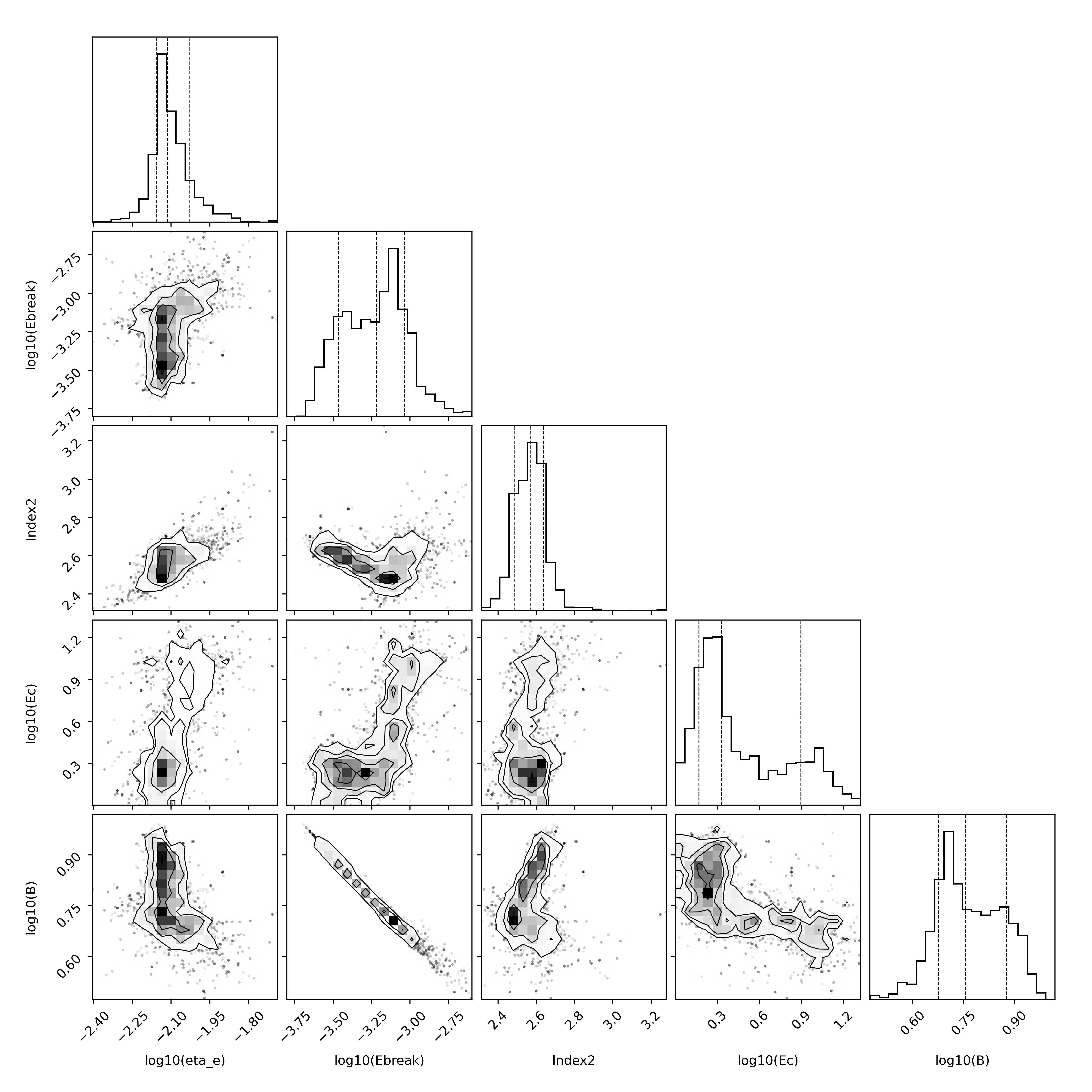}
        \caption{GRB~190114C corner plot for the wind medium.}
        \label{fig:corner_190114c_wind}
    \end{minipage}
    
\end{figure*}

\begin{figure*}
    \centering
    \begin{minipage}[t]{0.45\textwidth}
        \centering
        \includegraphics[width=\linewidth]{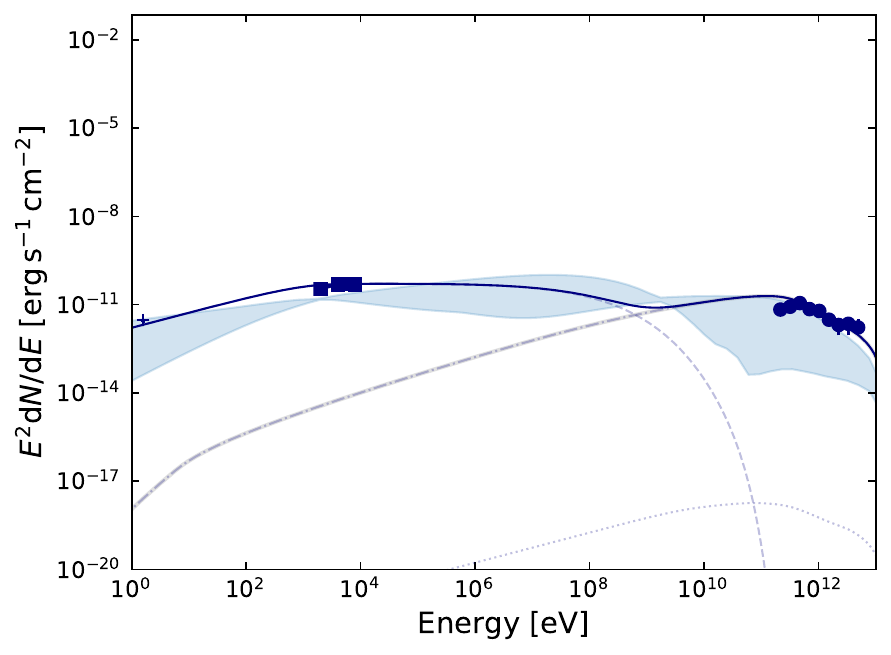}
        \caption{SED fitting of GRB~190829A in the ISM medium. The MCMC $1\sigma$-confidence band is shown over the model for the best-fit parameters are tabulated in Table \ref{tab:bestfit_params}.}
        \label{fig:mcmc_190829a_ism}
    \end{minipage}
    \hfill
    \begin{minipage}[t]{0.45\textwidth}
        \centering
        \includegraphics[width=\linewidth]{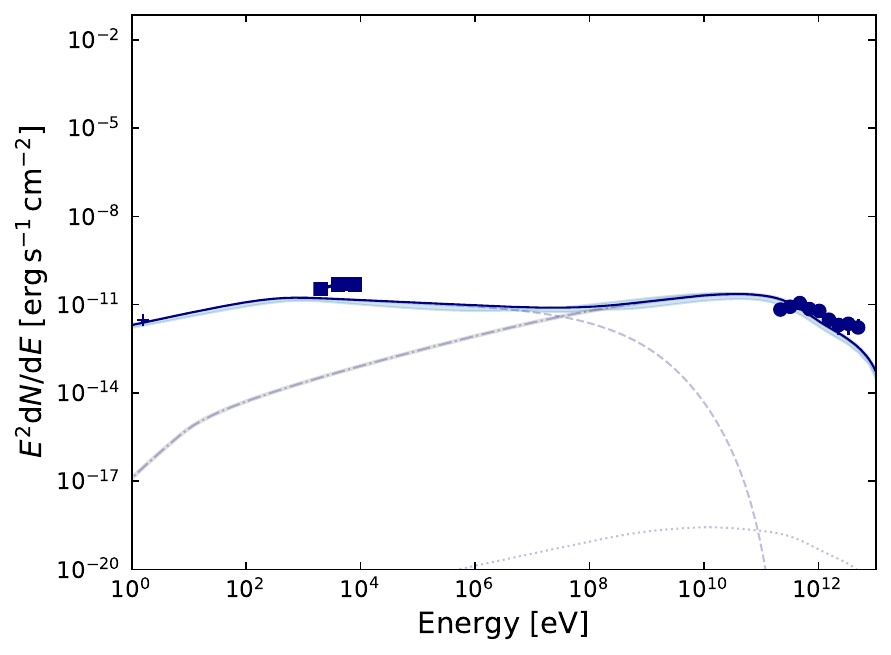}
        \caption{SED fitting of GRB~190829A for the wind medium. The MCMC $1\sigma$-confidence band is shown over the model for the best-fit parameters are tabulated in Table \ref{tab:bestfit_params}.}
        \label{fig:mcmc_190829a_wind}
    \end{minipage}
     
\end{figure*}

\begin{figure*}
    \centering
    \begin{minipage}[t]{0.48\textwidth}
        \centering
        \includegraphics[width=\linewidth]{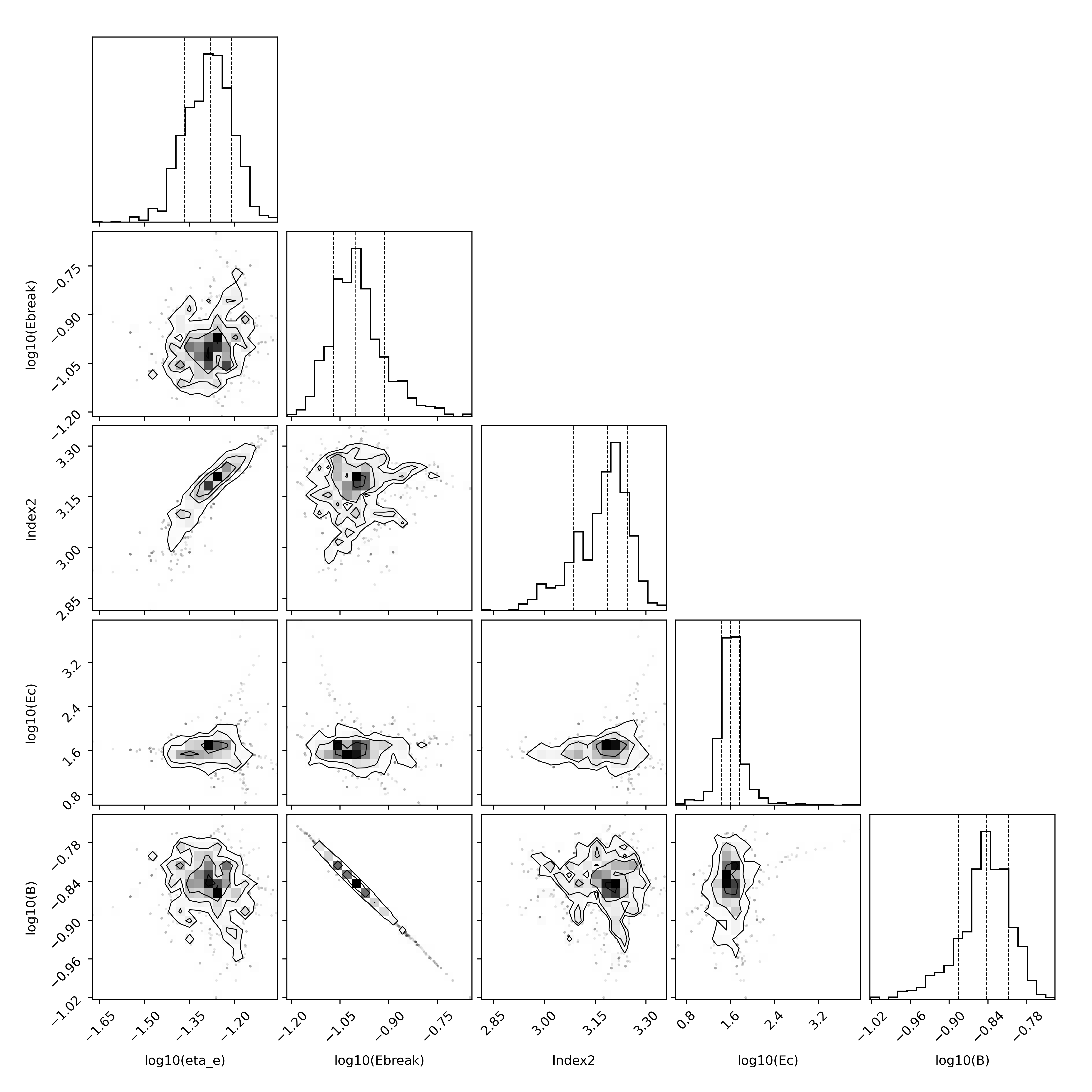}
        \caption{GRB~190829A corner plot for the ISM medium.}
        \label{fig:corner_190829a_ism}
    \end{minipage}
    \hspace{0.01\textwidth} 
    \begin{minipage}[t]{0.48\textwidth}
        \centering
        \includegraphics[width=\linewidth]{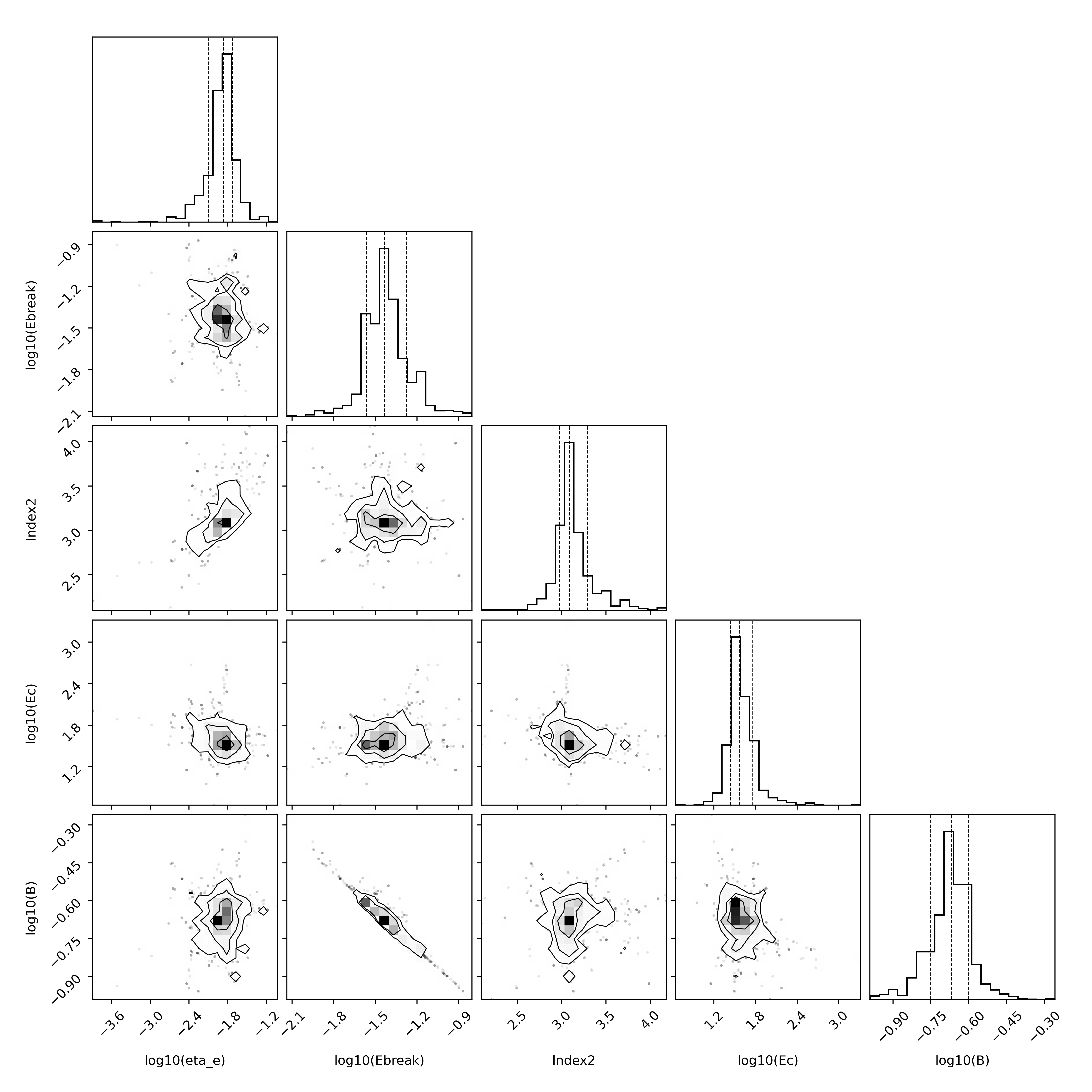}
        \caption{GRB~190829A corner plot for the wind medium.}
        \label{fig:corner_190829a_wind}
    \end{minipage}
    
\end{figure*}

\begin{figure*}
    \centering
    \begin{minipage}[t]{0.45\textwidth}
        \centering
        \includegraphics[width=\linewidth]{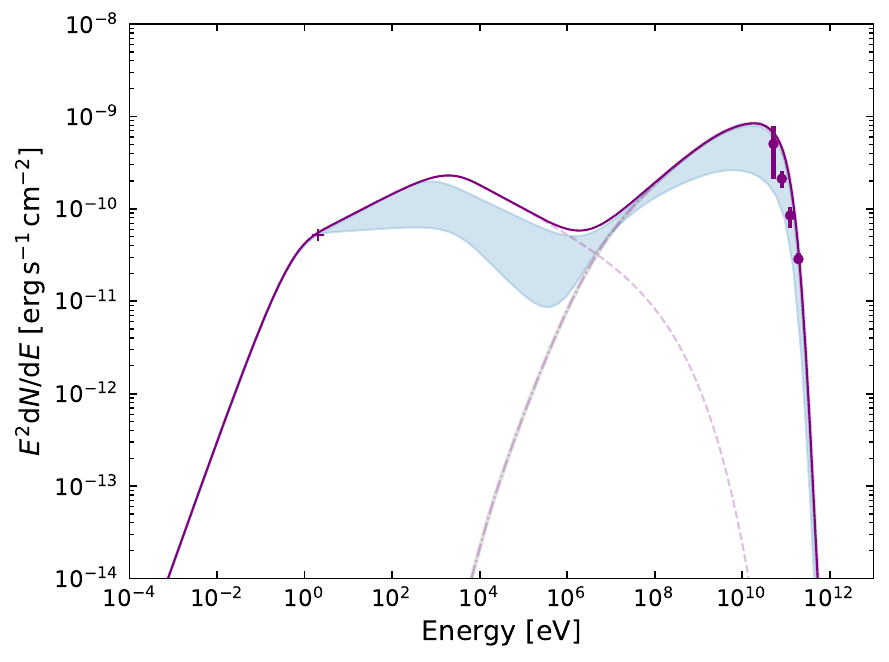}
        \caption{SED fitting of GRB~201216C in the ISM medium. The MCMC $1\sigma$-confidence band is shown over the model for the best-fit parameters are tabulated in Table \ref{tab:bestfit_params}.}
        \label{fig:mcmc_201216c_ism}
    \end{minipage}
    \hfill
    \begin{minipage}[t]{0.45\textwidth}
        \centering
        \includegraphics[width=\linewidth]{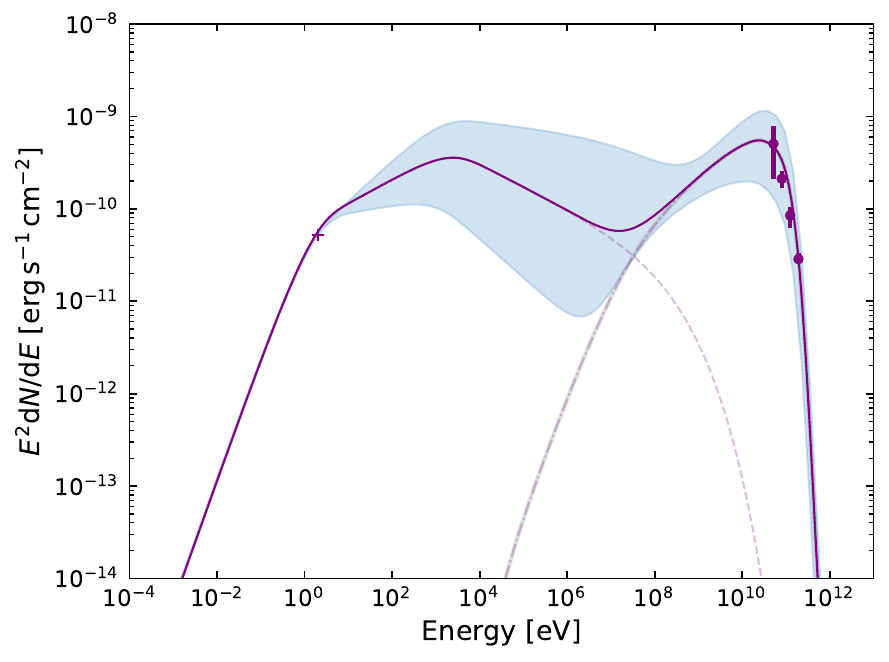}
        \caption{SED fitting of GRB~201216C for the wind medium. The MCMC $1\sigma$-confidence band is shown over the model for the best-fit parameters are tabulated in Table \ref{tab:bestfit_params}.}
        \label{fig:mcmc_201216c_wind}
    \end{minipage}
     
\end{figure*}

\begin{figure*}
    \centering
    \begin{minipage}[t]{0.48\textwidth}
        \centering
        \includegraphics[width=\linewidth]{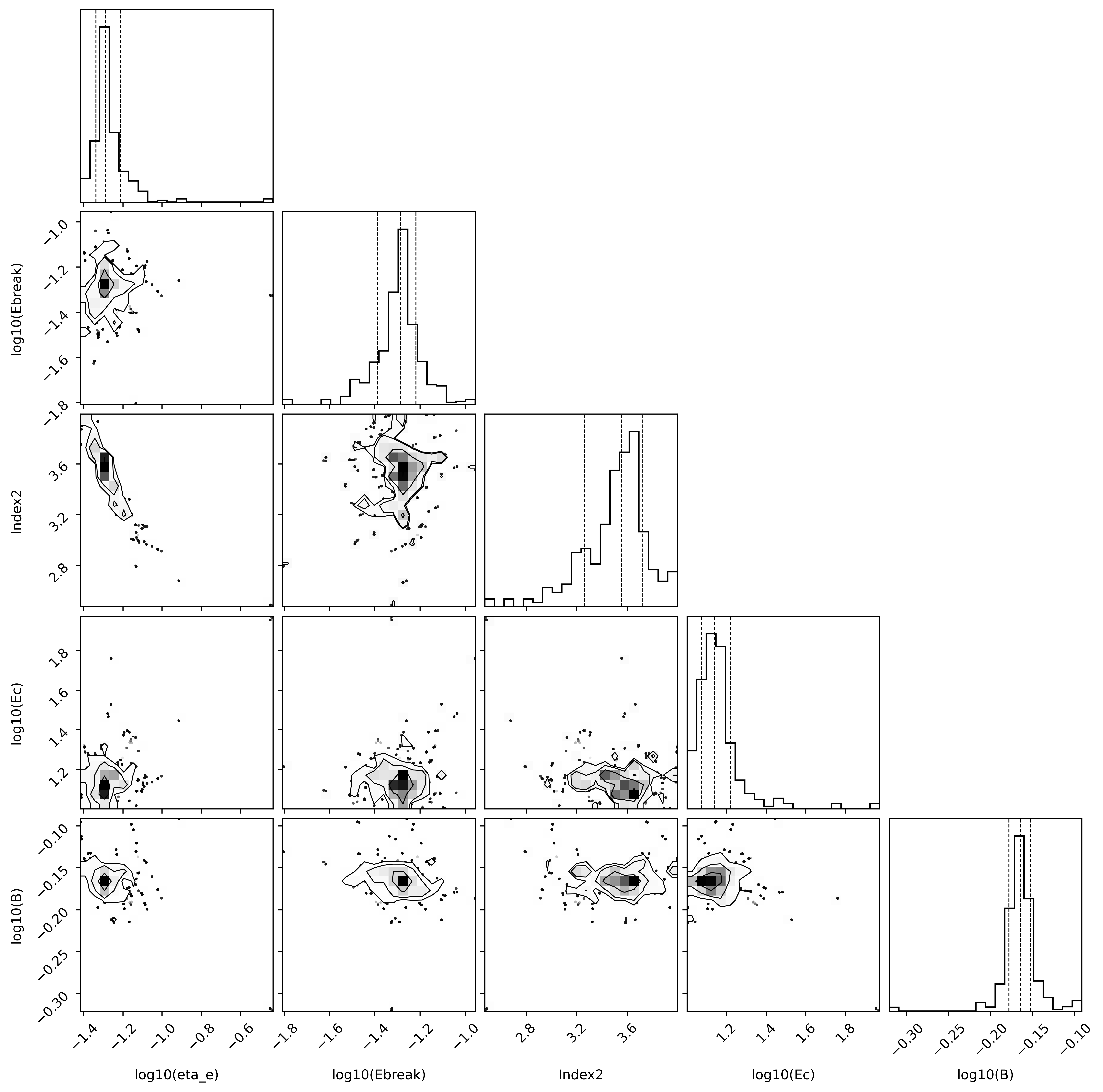}
        \caption{GRB~201216C corner plot for the ISM medium.}
        \label{fig:corner_201216c_ism}
    \end{minipage}
    \hspace{0.01\textwidth} 
    \begin{minipage}[t]{0.48\textwidth}
        \centering
        \includegraphics[width=\linewidth]{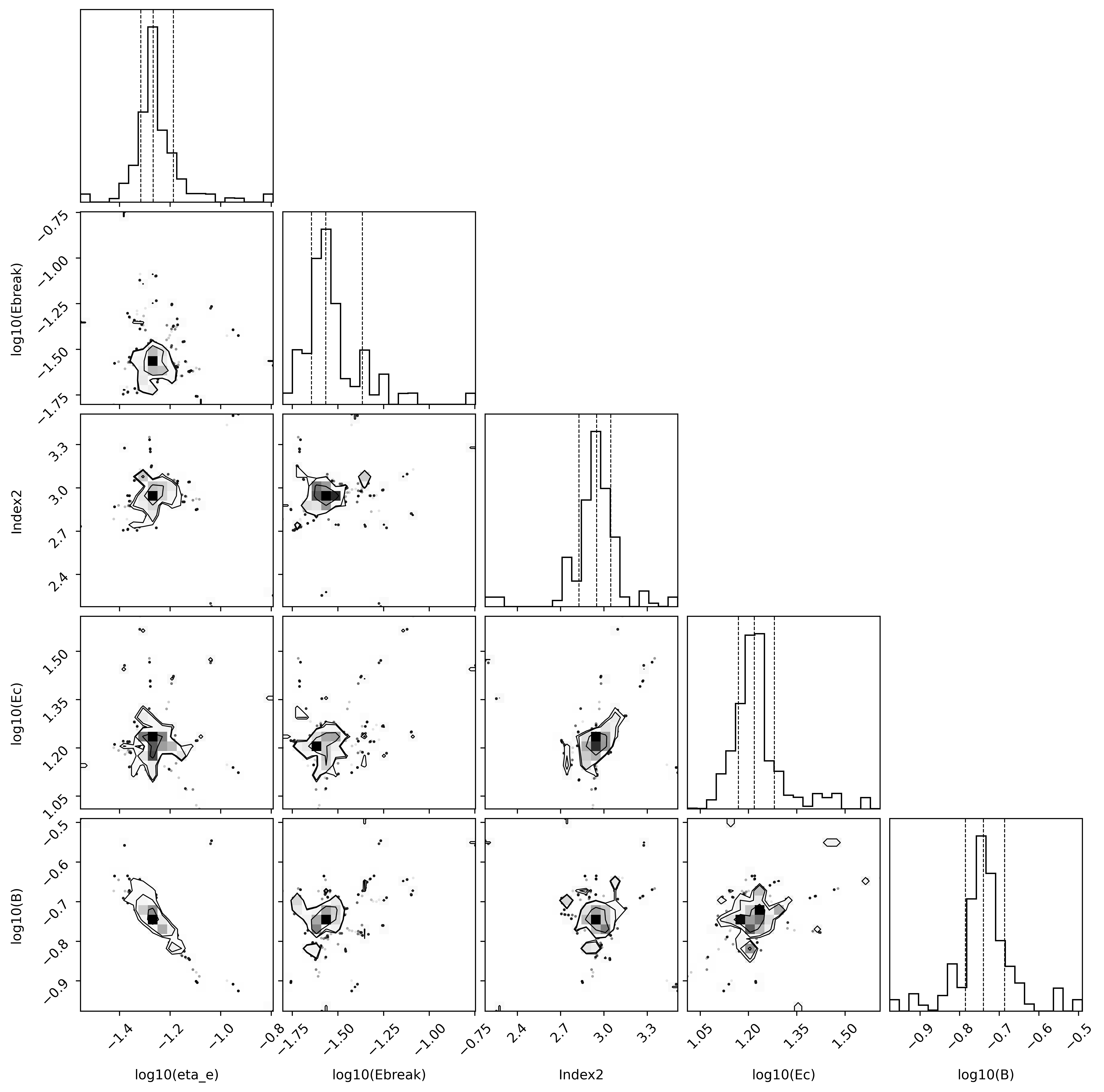}
        \caption{GRB~201216C corner plot for the wind medium.}
        \label{fig:corner_201216c_wind}
    \end{minipage}
    
\end{figure*}

\begin{figure*}
    \centering
    \begin{minipage}[t]{0.45\textwidth}
        \centering
        \includegraphics[width=\linewidth]{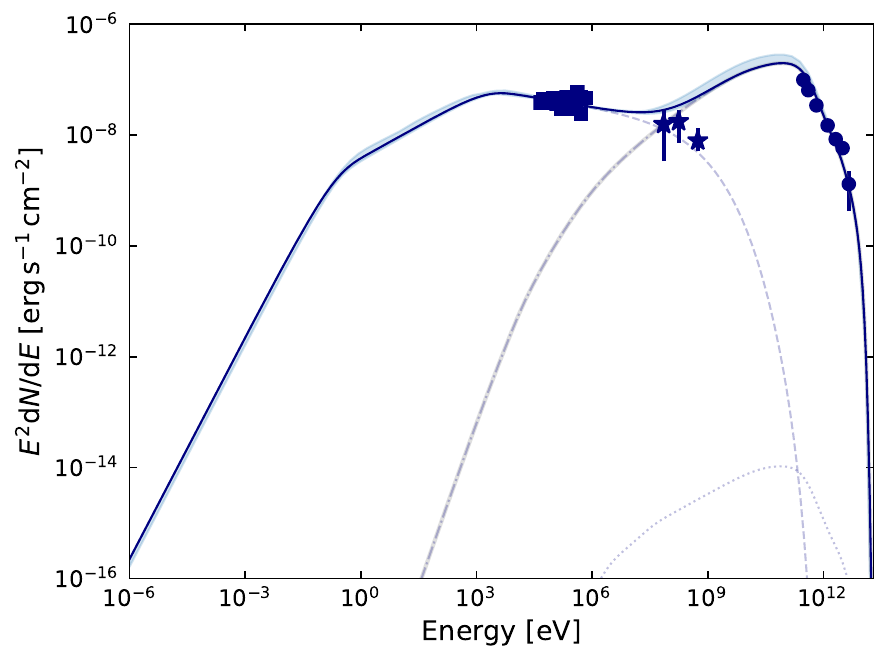}
        \caption{SED fitting of GRB~221009A in the ISM medium. The MCMC $1\sigma$-confidence band is shown over the model for the best-fit parameters are tabulated in Table \ref{tab:bestfit_params}.}
        \label{fig:mcmc_221009a_ism}
    \end{minipage}
    \hfill
    \begin{minipage}[t]{0.45\textwidth}
        \centering
        \includegraphics[width=\linewidth]{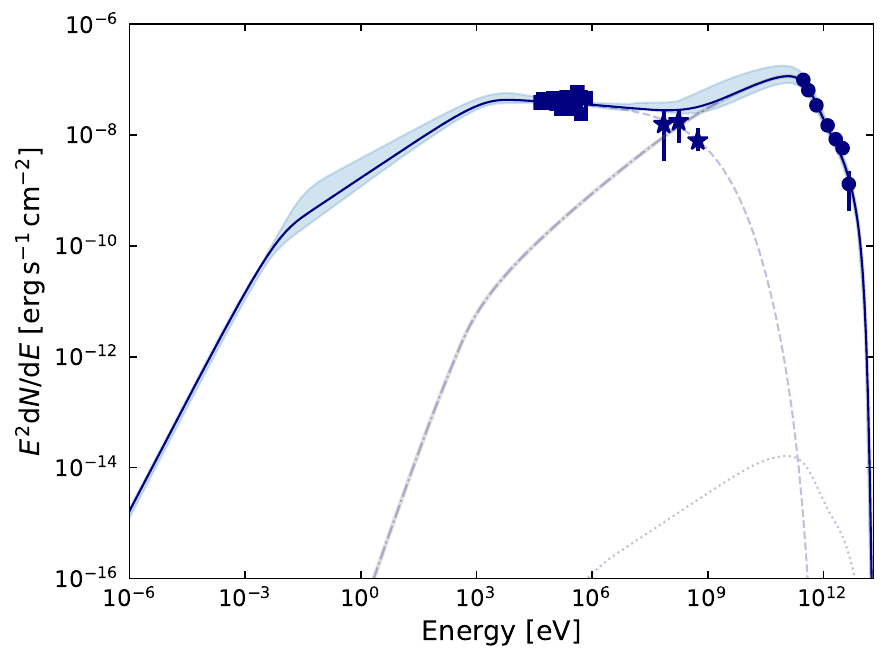}
        \caption{SED fitting of GRB~221009A for the wind medium. The MCMC $1\sigma$-confidence band is shown over the model for the best-fit parameters are tabulated in Table \ref{tab:bestfit_params}.}
        \label{fig:mcmc_221009a_wind}
    \end{minipage}
     
\end{figure*}

\begin{figure*}
    \centering
    \begin{minipage}[t]{0.48\textwidth}
        \centering
        \includegraphics[width=\linewidth]{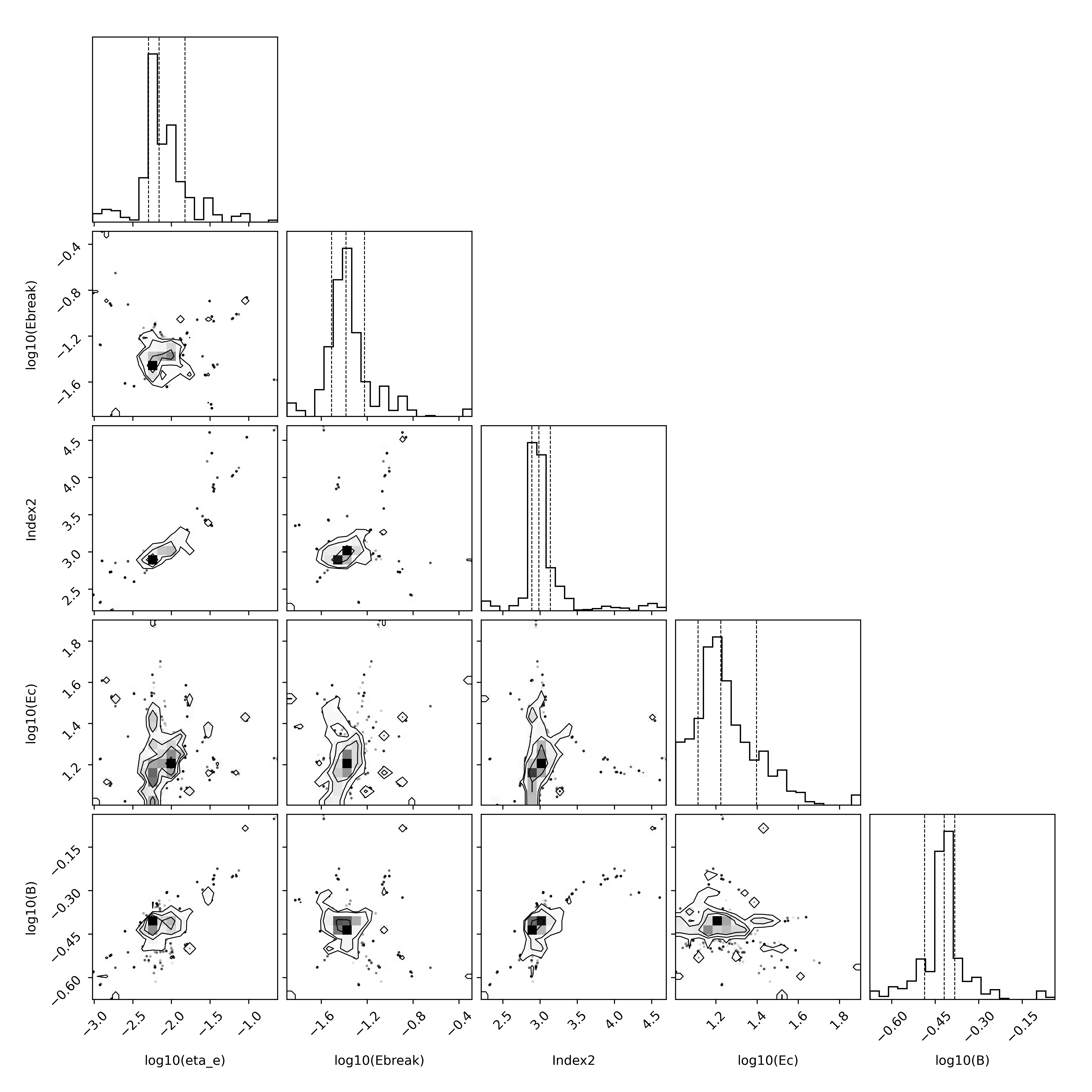}
        \caption{GRB~221009A corner plot for the ISM medium.}
        \label{fig:corner_221009a_ism}
    \end{minipage}
    \hspace{0.01\textwidth} 
    \begin{minipage}[t]{0.48\textwidth}
        \centering
        \includegraphics[width=\linewidth]{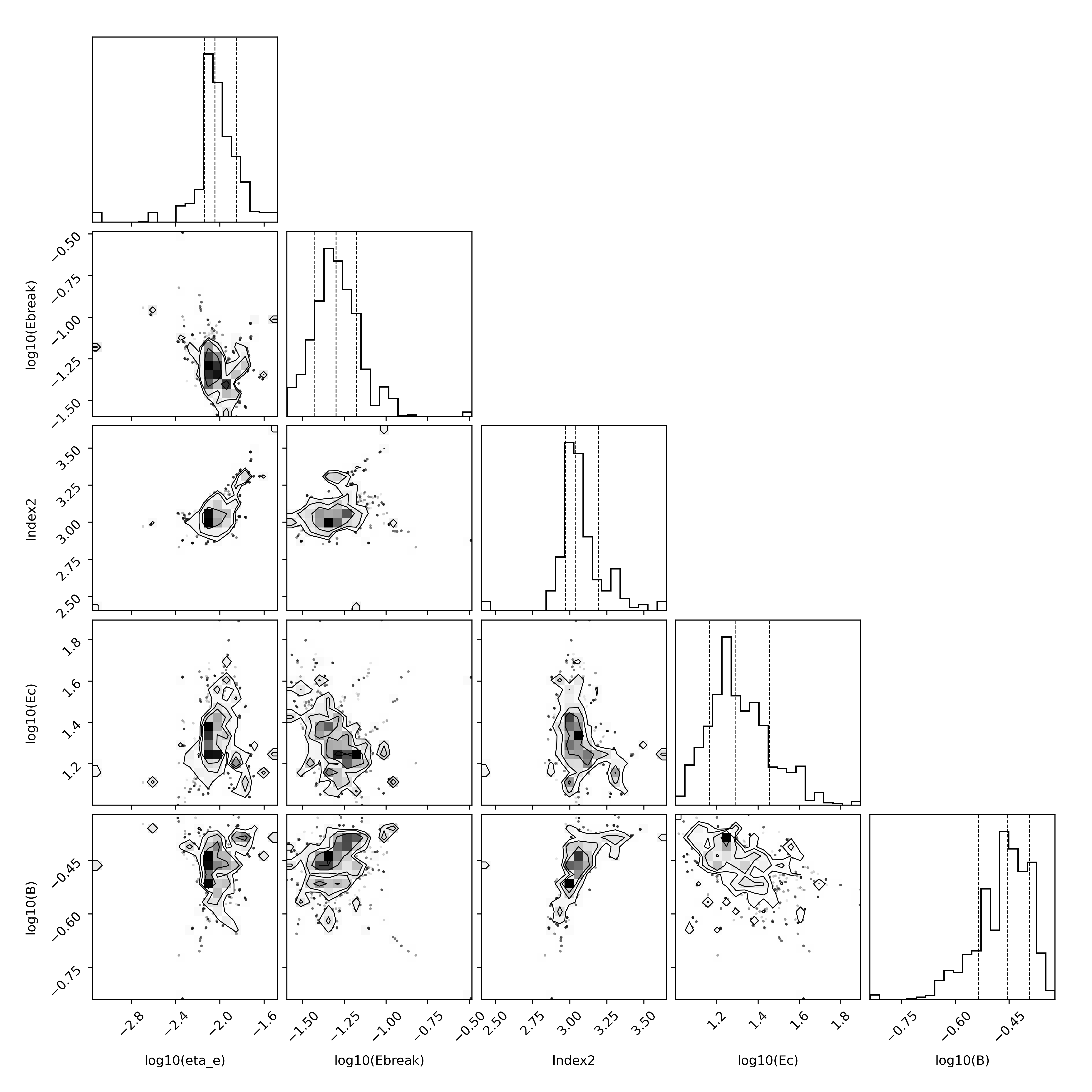}
        \caption{GRB~221009A corner plot for the wind medium.}
        \label{fig:corner_221009a_wind}
    \end{minipage}
    
\end{figure*}


\bsp	
\label{lastpage}
\end{document}